\documentclass[ reprint, superscriptaddress, amsmath,amssymb,prx, longbibliography]{revtex4-2}

\usepackage[normalem]{ulem}
\usepackage[dvipsnames]{xcolor}

\usepackage{amsmath}
\usepackage{amsfonts}
\usepackage{pgfplots}
\DeclareUnicodeCharacter{2212}{−}
\usepgfplotslibrary{groupplots,dateplot}
\usetikzlibrary{patterns,shapes.arrows}
\usepackage{bbm}
\usepackage{cancel}
\usepackage{braket}
\usepackage{dsfont}
\usepackage{enumerate}
\usepackage{url}
\begin{document}
\title{Crafting the dynamical structure of synchronization \\by harnessing bosonic multilevel cavity QED}
\author{Riccardo J. Valencia-Tortora} \email[]{Corresponding author: rvalenci@uni-mainz.de}
\affiliation{Institut f\"{u}r Physik, Johannes Gutenberg-Universit\"{a}t Mainz, D-55099 Mainz, Germany}    
\author{Shane P. Kelly}
\affiliation{Institut f\"{u}r Physik, Johannes Gutenberg-Universit\"{a}t Mainz, D-55099 Mainz, Germany}
\author{\\Tobias Donner}
\affiliation{
Institute for Quantum Electronics, Eidgen\"{o}ssische Technische Hochschule Z\"{u}rich, Otto-Stern-Weg 1, CH-8093 Zurich, Switzerland}
\author{Giovanna Morigi}
\affiliation{Theoretical Physics, Department of Physics, Saarland University, 66123 Saarbr\"{u}cken, Germany}
\author{Rosario Fazio}
\affiliation{The Abdus Salam International Center for Theoretical Physics (ICTP), I-34151 Trieste, Italy}
\affiliation{Dipartimento di Fisica, Universit\`a di Napoli Federico II, Monte S. Angelo, I-80126 Napoli, Italy}
\author{Jamir Marino}
\affiliation{Institut f\"{u}r Physik, Johannes Gutenberg-Universit\"{a}t Mainz, D-55099 Mainz, Germany}
\date{\today}%
\begin{abstract}
\noindent
Many-body cavity QED experiments are   established platforms to tailor  and control the collective responses of ensembles of atoms, interacting through one or more common photonic modes. The rich diversity of dynamical phases they can host, calls for a unified framework. 
Here we commence this program by showing that a cavity QED simulator assembled from $N$-levels bosonic atoms, can reproduce and extend the possible dynamical responses of collective observables occurring after a quench.
Specifically, by initializing the atoms in classical or quantum states, or by leveraging intra-levels {quantum correlations}, we craft on demand the entire synchronization/desynchronization dynamical crossover of an exchange model for $SU(N)$ spins. 
We quantitatively predict the onset of different dynamical responses by combining the Liouville-Arnold theorem on classical integrability with an ansatz for reducing the collective evolution to an effective few-body dynamics.
Among them, we discover a synchronized chaotic phase induced by {quantum correlations} and associated to a first order non-equilibrium transition in the Lyapunov exponent of collective atomic dynamics.    Our outreach includes extensions to other spin-exchange quantum simulators and a universal conjecture for the dynamical reduction of non-integrable all-to-all interacting systems.  
\end{abstract}
\maketitle
\section{Introduction}
Tailoring   light-matter interactions is at the root   of   numerous technological or experimental applications in quantum optics, and it has generated a persistent drive for better control of atoms and  photons since the advent of modern molecular  and  atomic physics.
For instance, the pursuit to create precision clocks and sensors has lead to the development of cavity QED systems in which a cold gas couples to few or several electromagnetic modes in an optical cavity~\cite{Baumann2010,ritsch2013cold,mivehvar2021cavity,PhysRevLett.91.203001,TanjiSuzuki2011,Bohnet2012}.
Such systems can be   brought out of equilibrium to generate reproducible many-body dynamics which show complex behavior including self-organization~\cite{brennecke2013real,leonard2017monitoring,PhysRevLett.91.203001,PhysRevLett.121.163601,PhysRevLett.120.223602,PhysRevLett.107.140402,PhysRevLett.115.230403,PhysRevX.8.011002,Nagy2006,Nagy2008} and dynamical phase transitions~\cite{klinder2015dynamical,PhysRevLett.115.230403,bakhtiari2015nonequilibrium,Norcia2018,muniz2020exploring,Bohnet2012,Zhiqiang:17,Baumann2010}, quantum squeezed and non-Gaussian entangled states~\cite{cox2016deterministic,PhysRevLett.121.070403,pedrozo2020entanglement,colombo2022time,Barontini2015,Hosten2016,PhysRevLett.104.073602}, time crystals~\cite{kessler2021observation,Kongkhambut2022,Dogra2019,Dreon2022}, and glassy dynamics~\cite{PhysRevX.8.011002,PhysRevLett.121.163601,guo2019sign,marsh2021enhancing}.
This   rich phenomenology  comes from a high degree of tunability in such systems, allowing control over local external fields, detunings between cavity mode and applied drive fields, the ability to couple multiple atomic levels to the cavity field~\cite{TanjiSuzuki2011,PhysRevLett.121.173602,PhysRevLett.127.253601,PhysRevA.97.043858,PhysRevLett.122.010405,PhysRevLett.121.163601,PhysRevX.11.041046,Marino2019}, and more recently the realization of programmable geometries for light-matter interactions~\cite{Periwal2021,PhysRevB.105.184305,PhysRevLett.129.050603}.\\
  
Recently, the theoretical and experimental investigation of multilevel cavity systems has  gathered increasing attention. 
Current progress includes dissipative state preparation of entangled dark states~\cite{PhysRevX.12.011054,PhysRevLett.128.153601,PhysRevA.84.053856}, multicriticality in generalized Dicke-type models~\cite{PhysRevA.104.043708,PhysRevA.101.063627}, incommensurate time
crystalline phases~\cite{PhysRevLett.127.253601,PhysRevA.104.063705,PhysRevA.100.053615}, correlated pair creations and phase-coherence protection via spin-exchange interactions~\cite{PhysRevLett.122.010405,PhysRevLett.125.060402,perlin2022engineering}, spin squeezing and atomic clock precision enhancement~\cite{PhysRevA.104.023710,PhysRevLett.109.173603,PhysRevX.8.021036}.
Yet, the quenched dynamics in multilevel cavity systems is widely unexplored and the few individual results lack an organizing principle.  

In this work, we propose a unifying framework for the dynamics after a quench of all-to-all connected multilevel systems.
We show that the flexible control endowed by bosonic multilevel atoms is sufficient to reproduce established dynamical phases and beyond.
We explain how the dynamical response can be crafted into these new and existing dynamical phases by introducing a reduction of  dynamics to a few-body effective classical evolution, valid regardless of the underlying  integrability of the model. 
Of particular note, we demonstrate how quantum {correlations} in the initial state can drive a transition between a regular and chaotic synchronized phases.

Our analysis extends the established phenomenology of the two-level Tavis-Cummings model with  local inhomogeneous fields.
This two-level model is integrable~\cite{kirton2018superradiant}, and allows for the emergent collective many-body dynamics to be exactly described through an effective few-body Hamiltonian~\cite{PhysRevA.91.033628,RevModPhys.76.643,richardson2002new,PhysRevLett.96.230403,PhysRevLett.93.160401,Gaudin1976,RICHARDSON1964221,PhysRevLett.96.097005,Yuzbashyan_2005,PhysRevB.72.220503,kelly2022resonant}. 
In particular, the few body model yields predictions for the dynamical responses of collective observables $S(t)$, such as the collective spin raising operator, given by the macroscopic sum of  several individual constituents~\cite{PhysRevA.91.033628,PhysRevLett.96.230403,PhysRevLett.93.160401,PhysRevLett.96.097005,Yuzbashyan_2005,PhysRevB.72.220503,kelly2022resonant,smale2019observation}. 
The resulting dynamical phases are best presented in terms of the possible synchronization between the local atomic degree of freedoms (spins-$1/2$)  which evolve with a frequency set by the competition of their local field and   collective photon-mediated interactions. 
In the desynchronized phase, which we call Phase-I as shorthand, all the spins evolve independently as a result of dominant classical dephasing processes imprinted by the local inhomogeneous fields, thus $S(t)$ relaxes to zero. 
In the synchronized phase, collective interactions lock the phase precession and we can distinguish three different scenarios in which $S(t)$ either relaxes to a stationary value (Phase-II), up to a phase of a Goldstone mode~\cite{PhysRevA.91.033628} associated to a global $U(1)$ symmetry, or its magnitude enters self-generated oscillatory dynamics, corresponding to a Higgs mode~\cite{PhysRevA.91.033628}, either periodic (Phase-III), or aperiodic (Phase-IV). While Phase-I and Phase-II describe relaxation to a steady state up to an irrelevant global phase, Phase-III and Phase-IV are instead examples of a self-generated oscillating synchronization phenomenon without an external driving force~\cite{PhysRevLett.115.163601,Zhu2019,Tucker2018,https://doi.org/10.48550/arxiv.2112.04509}.

\subsection{Summary of results}

In this work we   investigate   dynamics beyond two-level approximations  by considering $\mathcal{N}_a$ bosonic atoms, each hosting $N$ levels which realize $SU(N)$ spins. 
The additional structure due to the bosonic statistics allows us to naturally consider both classical and quantum initial states~(c.f. Sec.~\ref{sec_initial_states}).
Using this flexibility in the initial state, and also the tunability of Hamiltonian parameters, we show how to craft not only the dynamical responses present in the two-level integrable setup (from Phase-I up to Phase-IV), but also how to access a novel chaotic dynamical response. 
This chaotic response, which we refer to as Phase-IV${}^\star$, again has all atoms synchronized but with the dynamics of the average atomic coherences characterized by exponential sensitivity to initial conditions.
The self-generated chaotic Phase-IV${}^\star$ emerges from the interplay of initial {quantum correlations}, and the collective interactions mediated by the cavity field. It is therefore qualitatively different from chaos induced by other mechanisms as due to additional local interactions~\cite{PhysRevLett.120.130603} or external pump~\cite{PhysRevLett.123.053601,PhysRevA.101.023616,Dogra2019,PhysRevLett.122.193605,PhysRevA.95.023806}. 

\begin{figure}[t!]
\centering
\includegraphics[width=\linewidth]{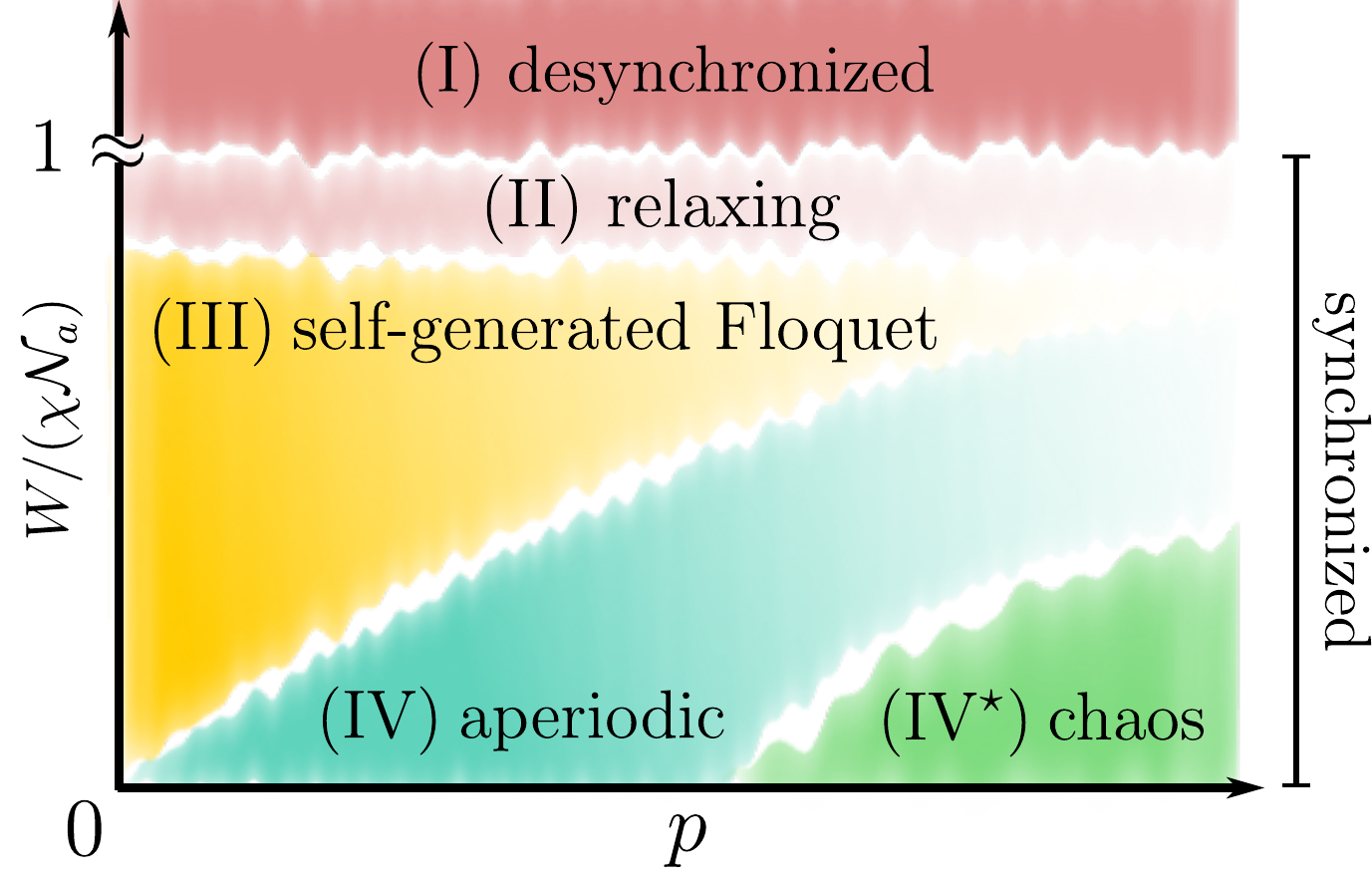}
\caption{
Cartoon of the possible dynamical responses of intra-level phase coherence in a photon-mediated spin-exchange model between $SU(3)$ spins, as a function of the degree of inhomogeneity of the local fields $W$, and of {quantum correlations} in the initial state parameterized by $p$. At $p=0$ each site is initialized in the same bosonic coherent state. For $p>0$, there are finite {quantum correlations} in the system. The parameter $p$ tunes from bosonic coherent states ($p=0$) to a multimode Schr\"{o}dinger cat state ($p>0$) initialized on each site. 
{The susceptibility of the dynamical response to quantum correlations is strictly linked to having $SU(N)$ spins with $N>2$, thus cannot be achieved considering two-level systems.}
Up to inhomogeneity $W/(\chi\mathcal{N}_a)\approx 1$, the system is in the synchronized phase. At larger inhomogeneities, the system enters in the desynchronized phase and all phase coherence is washed (Phase-I). In the synchronized phase, phase coherence relaxes asymptotically to a nonzero value up to a phase associated to a global $U(1)$ symmetry (Phase-II), or its magnitude enters a self-generated oscillatory dynamics, either periodic (Phase-III), or aperiodic (Phase-IV), as well as potentially chaotic (Phase-IV${}^\star$). In this last case dynamics are exponentially sensitive to changes in  initial conditions. 
\label{fig:cartoon}}
\end{figure}

In order to show how to craft and control these dynamical responses, we introduce a generalization of the reduction hypothesis used for two level systems.
Specifically, we propose that the different dynamical phases (Phase-I up to Phase-IV${}^*$) all correspond to a different effective few body Hamiltonian that depends on the global symmetries of the many body system, degree of inhomogeneity, $W$, number of atomic levels $N$, and degree of {quantum correlations} in the initial state, quantified by a parameter $p$ (cf.  Sec.~\ref{sec_chaos_induced_fluctuactions}). 
Then, by considering an appropriate classical limit arising in the limit of large system size~ (cf.  Sec.~\ref{sec_semiclassical_limit}), we apply the Liouville-Arnold theorem to the effective Hamiltonian to identify a correspondence between the dynamical phases and the effective Hamiltonians.
Using physical arguments for the nature of the effective Hamiltonian, we then predict how to tune between different dynamical responses.
The result is an intuitive control over the rich dynamical response possible in multilevel cavity QED.
See Fig.~\ref{fig:cartoon} for a cartoon of the different dynamical responses   for $N=3$ level atoms, using as a proxy the synchronized (or de-synchronized) evolution of the magnitude of the average intra-level coherences in the ensemble.

We conclude by discussing the potential universality of the reduction hypothesis.
In particular, we conjecture it applies not only for  state-of-the-art cavity QED experiments (cf. Sec.~\ref{sec_experimental_implementation}), but could find potential applications in other fields. Following  Refs.~\cite{PhysRevLett.126.173601,PRXQuantum.3.040324}, where   cavity QED  platforms  are proposed to model the dynamics  of  $s$-wave and $(p+ip)$-wave BCS superconductors, our results could find potential applications to lattice systems with local $SU(N)$ interactions, such as $SU(N)$ Hubbard models~\cite{PhysRevA.104.043316,PhysRevA.85.041604,PhysRevLett.109.205305,PhysRevA.89.043610}. Another possible outreach of our results could consist in noticing that the $N$ levels of the atoms could be used as a synthetic dimension, with the geometry fixed by the photon-mediated processes, as for instance in  a synthetic ladder system~\cite{https://doi.org/10.48550/arxiv.2204.06421,https://doi.org/10.48550/arxiv.2208.01896}.
Furthermore, since we consider bosonic systems, our results could potentially find applications in spinor Bose-Einstein condensates~\cite{RevModPhys.90.035005,rodriguez2021universal} or in molecules embedded in a cavity, where bosons could be identified as their vibrational modes~\cite{PhysRevLett.122.203602,campos2021generalization}.

\subsection{Organization of the manuscript}

The paper is organized as follows.
In Sec.~\ref{sec:introduction}, we introduce the model and initial states we investigate, and we discuss the cumulant expansion we use to capture quench dynamics.
In Sec.~\ref{sec_reduction_hypothesis}, we present the dynamical reduction hypothesis, and discuss the different classes of effective dynamics that can result from it.
In Sec.~\ref{sec_W0} we show that in the homogeneous limit our hypothesis is exact and demonstrate how local quantum {correlations} in $SU(3)$ atoms can induce a chaotic dynamical phase with finite Lyapunov exponent.
In Sec.~\ref{sec_Wfinite}, we show that the dynamical responses observed in the homogeneous limit   are robust against moderate inhomogeneity in the local fields, and we provide numerical evidences that an effective few-body Hamiltonian is able to capture the dynamical periodic response of collective observables in the three-level case.
We conclude this section with a discussion on the impact of inhomogeneity in the dynamical responses of the system.
In Sec.~\ref{sec_experimental_implementation} we propose an experimental implementation potentially accessible in state-of-the-art cavity QED systems.
\section{Preliminaries~\label{sec:introduction}}
\subsection{The model~\label{sec:model}}
We consider a system of $\mathcal{N}_a$ bosonic atoms interacting via a single photonic mode of a cavity.
{The atoms are cooled to the motional ground state and evenly distributed among $L$ different atomic ensembles labeled by a site index $j$. Within each site (ensemble), the atoms are indistinguishable and can occupy $N$ different atomic levels with energies that are site- and level-dependent. We consider the atoms sufficiently far apart for interatomic interactions to be negligible.}
The photon-matter interaction mediates atom number conserving processes where the absorption and/or the emission of a cavity photon results in an atom transitioning from level $n$ to levels $n\pm 1$ within the same site, with a rate generally dependent on the specific level $n$. 
The associated many-body light-matter Hamiltonian reads
\begin{equation}
\label{eq_H_SUn_wphoton}
\begin{split}
\hat{H} =& \omega_0 \hat{a}^\dagger \hat{a} + \sum_{j=1}^L\sum_{n=1}^N h_n^{(j)}\hat{b}_{n,j}^\dagger \hat{b}_{n,j}+\\
&+\sum_{j=1}^L \sum_{n=1}^{N-1} \Big[ g_{n} \left(\hat{b}_{n+1,j}^\dagger \hat{b}_{n,j}\hat{a} + h.c.\right)+\\
&\qquad\qquad+ \lambda_{n} \left(\hat{b}_{n+1,j}^\dagger \hat{b}_{n,j}\hat{a}^\dagger + h.c.\right)\Big],
\end{split}
\end{equation}
where $\hat{a}^{(\dagger)}$ is the bosonic annihilation (creation) operator of the cavity photon; $\hat{b}_{n,j}^{(\dagger)}$ is the bosonic annihilation (creation) operator on site $j\in[1,L]$ and level $n \in [1,N]$,  with energy splitting $h_n^{(j)}$; $g_n$ and $\lambda_n$ are the single-particle photon-matter couplings which controls rotating and co-rotating processes, respectively. Tuning $g_n$ and $\lambda_n$ enables us to pass from a generalized multilevel Dicke model, when $g_n,\lambda_n \neq 0$, to the multilevel Tavis-Cummings model, when $\lambda_n=0$. In our work, we consider dynamics on time scales where dissipative processes are sub-dominant compared to coherent evolution (cf. Sec.~\ref{sec_role_dissipation}). 

When the cavity is far detuned from the atomic transitions, the photon  does not actively participate in   dynamics of Eq.~\eqref{eq_H_SUn_wphoton} but instead  mediates virtual atom-atom interactions~\cite{CohenTannoudji1998}.
This occurs in the limit $\omega_0 \gg \max\{h_n^{(j)},g_n\sqrt{\mathcal{N}_a},\lambda_n\sqrt{\mathcal{N}_a}\}$, where the factor $\sqrt{\mathcal{N}_a}$ comes from the cooperative enhancement given by the $\mathcal{N}_a$ atoms~\cite{PhysRevLett.118.123602,Brennecke2007}.
The mediated interaction results in an effective atoms-only Hamiltonian of the form
\begin{equation}
\label{eq_H_sun_adiabatic}
\begin{split}
\hat{H} &=\sum_{j=1}^L\sum_{n=1}^N h_n^{(j)}\hat{\Sigma}_{n,n}^{(j)}+\\
 -&\sum_{m,n=1}^{N-1}\Big[\chi_{n,m}\hat{\Sigma}_{n+1,n} \hat{\Sigma}_{m,m+1} +\zeta_{n,m}  \hat{\Sigma}_{n,n+1} \hat{\Sigma}_{m+1,m}+\\
 &+\nu_{n,m} \hat{\Sigma}_{n+1,n} \hat{\Sigma}_{m+1,m} + \nu_{m,n}\hat{\Sigma}_{n,n+1} \hat{\Sigma}_{m,m+1}\Big],
\end{split}
\end{equation}
where $\chi_{n,m} \equiv g_n g_m /\omega_0$; $\zeta_{n,m}\equiv \lambda_n \lambda_m/\omega_0$; $\nu_{n,m} \equiv \lambda_n g_m /\omega_0$. For convenience, we have written the Hamiltonian in Eq.~\eqref{eq_H_sun_adiabatic} as a function of the operators 
\begin{align}
\hat{\Sigma}_{n,m}^{(j)}&=\hat{b}_{n,j}^\dagger \hat{b}_{m,j},\\
\label{eq_collective_operators}
\hat{\Sigma}_{n,m} &= \sum_{j=1}^L \hat{\Sigma}_{n,m}^{(j)}.
\end{align}
The operators $\{\hat{\Sigma}_{n,m}^{(j)}\}$ are generators of the $SU(N)$ group~\cite{auerbach2012interacting,zhang2021classical} and they obey the commutation relations $[\hat{\Sigma}_{n,m}^{(i)},\hat{\Sigma}_{k,l}^{(j)}]=\delta_{i,j}(\hat{\Sigma}_{n,l}^{(i)}\delta_{m,k}-\hat{\Sigma}_{k,m}^{(j)}\delta_{n,l})$, and $(\hat{\Sigma}_{n,m}^{(j)})^\dagger = \hat{\Sigma}_{m,n}^{(j)}$. 

The regime we are mostly interested in is $\nu_{n,m}=\zeta_{n,m}=0$, which translates to $\lambda_n=0$.
In this limit, the Hamiltonian in Eq.~\eqref{eq_H_sun_adiabatic} turns into a spin-exchange interaction Hamiltonian {between $SU(N)$ spins} with   rates $\{\chi_{n,m}\}$ and inhomogeneous fields, $h^{(j)}_n$.
In the following, we set the collective spin-exchange rate $\chi\mathcal{N}_a =\mathcal{N}_a \sum_{n=1}^{N-1}\chi_{n,n}$ as our energy scale, such that the time-scales of our results are independent of the number of atoms $\mathcal{N}_a$ in the system. An implementation of the spin exchange model in Eq.~\eqref{eq_H_sun_adiabatic} is offered in Sec.~\ref{sec_experimental_implementation}.

Below, we consider both situations when the energies of the atomic levels are homogenous and when they are inhomogenous.
In the latter situation, we expect our results to hold for various forms of inhomogenities, but we will in particular focus on the situations when the atomic levels on each site are in an evenly spaced ladder configuration with spacing $\Delta h_j \equiv (h_{n+1}^{(j)}-h_n^{(j)})$ sampled from a box distribution with zero average and width $W$.
In this case, the Hamiltonian is spatially homogeneous for $W=0$, and spatially inhomogeneous for $W>0$.
At $W=0$ we can make precise predictions of the dynamical responses as a function of the features of the initial state and multilevel structure. Then, we show numerically their robustness against many-body dynamics due to inhomogenities ($W>0$), in a fashion reminiscent of a synchronization phenomenon.

Given an evenly spaced ladder configuration within each site,  the Hamiltonians in Eq.~\eqref{eq_H_SUn_wphoton} and Eq.~\eqref{eq_H_sun_adiabatic} can, for certain values of the couplings $g_n$ and $\lambda_n$, be written in terms of the generators of a subgroup of $SU(N)$.
For instance, in the $N=3$ level case, if $g_n=g$ and $\lambda_n=\lambda$, the Hamiltonian can be written as a function of the generators of a $SU(2)$ subgroup of $SU(3)$.
Specifically, only the $SU(2)$ operators $\hat{S}_j^- = \sqrt{2}(\hat{\Sigma}_{1,2}^{(j)}+\hat{\Sigma}_{2,3}^{(j)})$, $\hat{S}_j^+ = (\hat{S}_j^-)^\dagger$, and $\hat{S}_j^z = (\hat{\Sigma}_{3,3}^{(j)}-\hat{\Sigma}_{1,1}^{(j)})$ are required to represent the Hamiltonian, and as a consequence, the dynamics can be more simply described by the dynamics of these $SU(2)$ spins.
For instance, we recover the spin-1 Dicke model for $\lambda=g$ and the spin-1 Tavis-Cummings model for $\lambda=0$ in Eq.~\eqref{eq_H_SUn_wphoton}. 
Since we aim to explore the impact of genuine interactions between $SU(N)$ spins, we fix $g_n$ and $\lambda_n$ such that the dynamics cannot be restricted to a subgroup of $SU(N)$, if not otherwise specified. An important exception is  the three-level case, where   the system can enter in a chaotic phase upon passing from interactions between $SU(2)$ to $SU(3)$ spins (see Sec.~\ref{sec_chaos_induced_fluctuactions}).
We highlight that while the interactions considered lead to nontrivial effects in the $SU(N)$ degrees of freedom, they are not $SU(N)$-symmetric. 

\subsection{Mean field limit \label{sec_semiclassical_limit}}
Given a generic interacting Hamiltonian, the dynamics of any $n$-point correlation function depends on higher order correlation functions -- a structure known as the BBGKY hierarchy~\cite{huang2009introduction}.
In fully connected systems, as in our case, the hierarchy can be efficiently truncated
starting from separable states, or in other words, from a Gutzwiller-type ansatz~\cite{PhysRevLett.10.159}
\begin{equation}
\label{eq_generic_state}
|\Psi\rangle = \otimes_{j=1}^L |\psi_{j}\rangle \otimes |\alpha\rangle,
\end{equation}
where $|\psi_j\rangle$ is a generic state {on} the $j$-th atom, and $|\alpha\rangle$ is a bosonic coherent state describing the cavity field.
Given   $|\Psi\rangle$ in Eq.~\eqref{eq_generic_state}{,} 
the hierarchy can be truncated as
$\langle \hat{\Sigma}_{n,m}^{(j)}\hat{a}\rangle =\langle \hat{\Sigma}_{n,m}^{(j)}\rangle \langle \hat{a}\rangle$ and $\langle \hat{\Sigma}_{n,m}^{(j)}\hat{\Sigma}_{r,s}\rangle = \langle \hat{\Sigma}_{n,m}^{(j)}\rangle \langle \hat{\Sigma}_{r,s}\rangle$ up to $1/L$ corrections~\cite{PhysRevLett.118.123602,kirton2018superradiant,kirton2019introduction,PhysRevLett.126.230601,fiorelli2023meanfield}.
Here and from now on, we assume all expectation values are taken with respect to the state $\ket{\Psi}$, i.e. $\langle \hat{o}(t)\rangle \equiv \langle \Psi|\hat{o}(t)|\Psi\rangle$.
In the limit $L\to\infty$ no additional quantum correlations build up in time, hence the equation of motions of one-point and two-points correlation functions are exactly closed at all times and the state $|\Psi\rangle$ remains an exact ansatz of the many-body state. 

Combining the large $L$ limit and the nature of the interaction in the Hamiltonian, the dynamics of $\langle\hat{\Sigma}^{(j)}\rangle$  and $\langle \hat{a}\rangle$ can be accordingly obtained in the mean field limit of the Hamiltonians in Eq.~\eqref{eq_H_SUn_wphoton} and Eq.~\eqref{eq_H_sun_adiabatic}. This is achieved replacing the operators $\hat{\Sigma}_{n,m}^{(j)}$ and $\hat{a}^{(\dagger)}$ by classical $SU(N)$ spins and photon amplitude given by  
\begin{equation}
\label{eq_classical_spins}
\begin{split}
\Sigma_{n,m}^{(j)} &= \langle \hat{\Sigma}_{n,m}^{(j)}\rangle /(\mathcal{N}_a/L),\\
a &= \langle \hat{a} \rangle / \sqrt{\mathcal{N}_a},
\end{split}
\end{equation}
with $\mathcal{N}_a/L$ the average number of bosonic excitations per site and by substituting the commutators with Poisson brackets. The same dynamics can be obtained starting from the Heisenberg equation of motions 
and then taking the expectation value on the state $|\Psi\rangle$ in Eq.~\eqref{eq_generic_state}~\cite{zhang2021classical} truncating the hierarchy as discussed above. 

The hierarchy can be further truncated at first order in the bosonic operators if the one-body reduced density matrix $\Sigma^{(j)}$, with matrix elements $\Sigma_{n,m}^{(j)}$, is pure ($\text{Tr}[(\Sigma^{(j)})^2]=1$), namely {there are no quantum correlations on a given site $j$}. 
For instance, if the state $|\psi_j\rangle$ in Eq.~\eqref{eq_generic_state} is a bosonic coherent state on each level of site $j$, the matrix $\Sigma^{(j)}$ is pure and straightforwardly factorized as $\Sigma_{n,m}^{(j)} = \langle \hat{b}_{n,j}^\dagger \rangle \langle \hat{b}_{m,j}\rangle$. The truncation at first order in the bosonic operators well approximates the full dynamics up to corrections which are suppressed~\cite{Sciolla2011} in both the number of sites $L$ and the occupation on each site $\mathcal{N}_a/L$. 
Therefore, in the limit $\mathcal{N}_a\to \infty$, the hierarchy is exactly truncated at first order in the bosonic amplitudes $\langle\hat{b}_{n,j}^{(\dagger)}\rangle$ and $\langle  \hat{a}\rangle$, at all times. In this limit, their dynamics  
can be equivalently obtained in the classical limit of the Hamiltonians in Eq.~\eqref{eq_H_SUn_wphoton} and Eq.~\eqref{eq_H_sun_adiabatic} by replacing the bosonic operators $\hat{b}_{n,j}^{(\dagger)}$ and $\hat{a}$ by the classical fields
\begin{equation}
\label{eq_classical_bosonic_limit}
\begin{split}
b_{n,j} &= \langle\hat{b}_{n,j}\rangle/\sqrt{\mathcal{N}_a/L},\\
a &= \langle \hat{a}\rangle / \sqrt{\mathcal{N}_a},
\end{split}
\end{equation}
and replacing commutators with Poisson brackets.

{In the following sections we will investigate  the collective dynamical response of multilevel atoms in both mean field limits. We will show that the dynamical response could be highly susceptible to quantum correlations in the multilevel atom case, while it is insensitive in the two level case.}

\subsection{Initial states \label{sec_initial_states}}
In this work we derive general results which can be applied to any state of the form given in Eq.~\eqref{eq_generic_state}.
As discussed in Sec.~\ref{sec_semiclassical_limit} we distinguish two different classical limits, arising in the large $L$ limit, corresponding to the  one-body reduced density matrix $\Sigma^{(j)}$ on site $j$   being pure or mixed, respectively. 
For the sake of concreteness we now present a few states corresponding to the two cases discussed above.
The first two states are a bosonic coherent state and a $SU(N)$ spin-coherent state, both having {no quantum correlations and a one-body reduced density matrix that is pure}. While the other is a multimode Schr\"{odinger} cat state, whose one-body reduced density matrix on a given site is mixed reflecting {the presence of quantum correlations}.

\subsubsection{Coherent states}

The most general bosonic coherent state $|\psi_j\rangle$ on   a given site $j$ reads
\begin{equation}
\label{eq_bosonic_coherent_state}
\begin{split}
|\psi_j\rangle &=\exp\left(\boldsymbol{\gamma}_j \cdot \mathbf{\hat{b}_j^\dagger} -h.c.\right)|0\rangle \equiv |\widetilde{\gamma}_j\rangle,\\
\boldsymbol{\gamma}_j &\equiv (\gamma_{1,j},\gamma_{2,j},\dots , \gamma_{N,j}),\\
\mathbf{\hat{b}_j^\dagger}&\equiv(\hat{b}_{1,j}^\dagger,\hat{b}_{2,j}^\dagger,\dots,\hat{b}_{N,j}^\dagger),
\end{split}
\end{equation}
with $\gamma_{n,j} \in \mathbb{C}$ the amplitude of the bosonic coherent state on the $n$-th level and site $j$, {so that the average number of particles per site is $\sum_{n=1}^N |\gamma_{n,j}|^2=\mathcal{N}_a/L$. We highlight that the state in Eq.~\eqref{eq_bosonic_coherent_state} does not have an exact number of particles. Nonetheless, since the fluctuations of the number of particles are subleading with respect to the mean in the limit we consider ($\mathcal{N}_a/L \to \infty$), the mean field treatment is unaffected.}
Such a state has a pure single particle reduced density matrix, and will have an evolution captured by a mean field limit characterized by the classical variables $b_{n,j}$ and $a$.

\subsubsection{$SU(N)$ spin-coherent states \label{sec_SUN_spin_coherent_state}}
The second example of state with   pure one-body reduced density matrix is given by the  superposition: $|\psi_j\rangle = \sum_{n=1}^N \gamma_{n,j}\hat{b}_{n,j}^\dagger|0\rangle$, which has one excitation per site.
Once again, in this case, the mean field limit applies. Furthermore, the choice to truncate to one particle per site is insensitive  of  particles' statistics: either a fermion or boson could be the single particle occupying the site,  as we further elaborate in {the} concluding  section,   Sec.~\ref{sec_extension_results_to_other_systems}.
Such a state is the single particle limit of the more general $\mathcal{N}_a/L$ particle $SU(N)$ spin-coherent state~\cite{auerbach2012interacting} defined by
\begin{equation}
    \label{eq_spin_coherent_state}
    |\psi_j\rangle = \frac{1}{\sqrt{(\mathcal{N}_a/L)!}}\left(\sum_{n=1}^N \gamma_{n,j}\hat{b}_{n,j}^\dagger\right)^{\mathcal{N}_a/L}|0\rangle,
\end{equation}
which again has a pure one-body reduced density matrix reflecting a lack of {quantum correlations}. Thus, the dynamics of the classical variables $b_{n,j}$ and $a$ perfectly describe the dynamics of both the bosonic and $SU(N)$ spin coherent states in the limit of a large number of {bosons $\mathcal{N}_a$}.
Below we will present numerical results simulating these classical dynamics; they can be interpreted as describing the evolution of either of these two states. For the sake of simplicity, we will explicitly refer to these states as coherent states. 

\subsubsection{Schr\"{o}dinger cat states}
To consider a state in which the full two point correlations of the bosons, $\Sigma^{(j)}_{n,m}$, must be considered, {we add quantum correlations on} site $j$.
This ensures that the one body reduced density matrix is not pure and cannot be written in the mean field approximation, $\Sigma^{(j)}_{n,m}\neq b^*_{n,j}b_{m,j}$.
As an example, we consider a state where each site is initialized in a `multimode Schr\"{o}dinger cat state'~\cite{PRXQuantum.3.010301,PhysRevLett.123.260403}, which are the multimode generalization of `entangled coherent states'~\cite{PhysRevA.45.6811,Sanders2012,dodonov2002nonclassical}, given by the superposition of two bosonic coherent states $|\widetilde{\gamma}^{(m)}\rangle$ {with average occupation $\mathcal{N}_a/L$}, defined in Eq.~\eqref{eq_bosonic_coherent_state}, with $m = \{1,2\}$
\begin{equation}
\label{eq_generic_entangled_state}
|\psi_j\rangle = \frac{1}{\mathcal{D}}\left( |\widetilde{\gamma}_j^{(1)}\rangle +  |\widetilde{\gamma}_j^{(2)}\rangle\right).
\end{equation}
Here $\mathcal{D}$ is a normalization constant.  
If $|\langle \widetilde{\gamma}_j^{(1)}|\widetilde{\gamma}_j^{(2)}\rangle|= 1$ the state in Eq.~\eqref{eq_generic_entangled_state} reduces to the one in Eq.~\eqref{eq_bosonic_coherent_state}. 
Instead, if $|\langle \widetilde{\gamma}_j^{(1)}|\widetilde{\gamma}_j^{(2)}\rangle|< 1$, the one-body reduced density matrix is mixed, {reflecting the presence of quantum correlations on site $j$ ($\langle \hat{b}_{n,j}^{\dagger} \hat{b}_{m,j}\rangle_c \equiv \langle \hat{b}_{n,j}^{\dagger} \hat{b}_{m,j}\rangle - \langle \hat{b}_{n,j}^{\dagger}\rangle \langle \hat{b}_{m,j}\rangle \neq 0$).
We anticipate that the collective dynamical response could be highly susceptible to quantum {correlations} in the multilevel atom case, while they do not play a role in the two-level case. As an instance, we discover the onset of chaos as $|\langle \hat{b}_{n,j}^{\dagger} \hat{b}_{m,j}\rangle_c|$ increases in the $N=3$ levels case (cf. Sec.~\ref{sec_chaos_induced_fluctuactions}).}
We highlight that   quantum features of the state can only enter in   initial conditions since   dynamics are incapable of building   quantum correlations  in the mean field limit   (cf. Sec.~\ref{sec_semiclassical_limit}). 
\section{Classification of dynamical responses \label{sec_reduction_hypothesis}}
The main purpose of this work is to investigate and classify the dynamical response of collective observables in multilevel cavity QED systems in the long-time limit.
Specifically, we investigate the dynamics of the magnitude of the intra-level average coherences, defined as $|\sum_{j=1}^L \Sigma_{n,m}^{(j)}|/\mathcal{N}_a$ (for $n\neq m$).
To this end, we formulate the dynamical reduction hypothesis, which generalizes a similar procedure used for the integrable $SU(2)$ limits of Eq.~\eqref{eq_H_SUn_wphoton} and Eq.~\eqref{eq_H_sun_adiabatic}.
The hypothesis     conjectures that the dynamics of collective observables can be captured by the Hamiltonian dynamics of a few effective collective degrees of freedom (DOFs). In the integrable case, the effective Hamiltonian has been used to quantitatively predict the dynamical responses observed, which include relaxation and persistent oscillations either periodic or aperiodic ~\cite{kelly2022resonant,PhysRevA.91.033628,RevModPhys.76.643,PhysRevB.99.054520,richardson2002new,PhysRevLett.96.230403,PhysRevLett.93.160401,Gaudin1976,RICHARDSON1964221,PhysRevLett.96.097005,Yuzbashyan_2005,PhysRevB.72.220503}. Despite   lack of integrability,  
we still obtain in our case not simply relaxation, but also the persistent oscillatory  responses present in the integrable case, together with the possibility to develop  chaos~(see Fig.~\ref{fig_dynamics_homogeneous_initial_state_W0} for example)~\cite{PhysRevB.99.054520,dong2015dynamical,PhysRevB.104.104505}. Due to the generic non-integrable nature of multilevel systems an exact procedure for extracting  the effective model is not available (see Ref.~\cite{perlin2022engineering}, where the authors have attempted to extend the technique of the $SU(2)$ case to a $SU(N)$-symmetric interacting spin system). 

Here, we conjecture that, if an effective model exists, it is solely determined by the symmetries of the microscopic many-body problem and the relevant effective DOFs. Once the effective Hamiltonian is fixed, we show that the classification of dynamical responses follows from the combination of 1) the Liouville-Arnold theorem~\cite{goldstein2002classical}, which sets the criteria to distinguish a regular from an irregular (likely chaotic) regime, and 2) of the number of symmetries under which a given observable of interest is not invariant. 
Analogously to the integrable cased mentioned above, we offer a classification of  dynamical responses richer than the mere  distinction between desynchronization and synchronization.

\subsection{Dynamical Reduction Hypothesis \label{sec_universal_reduction_hypothesis}}
\begin{figure}[t!]
\centering
\includegraphics[width=\linewidth]{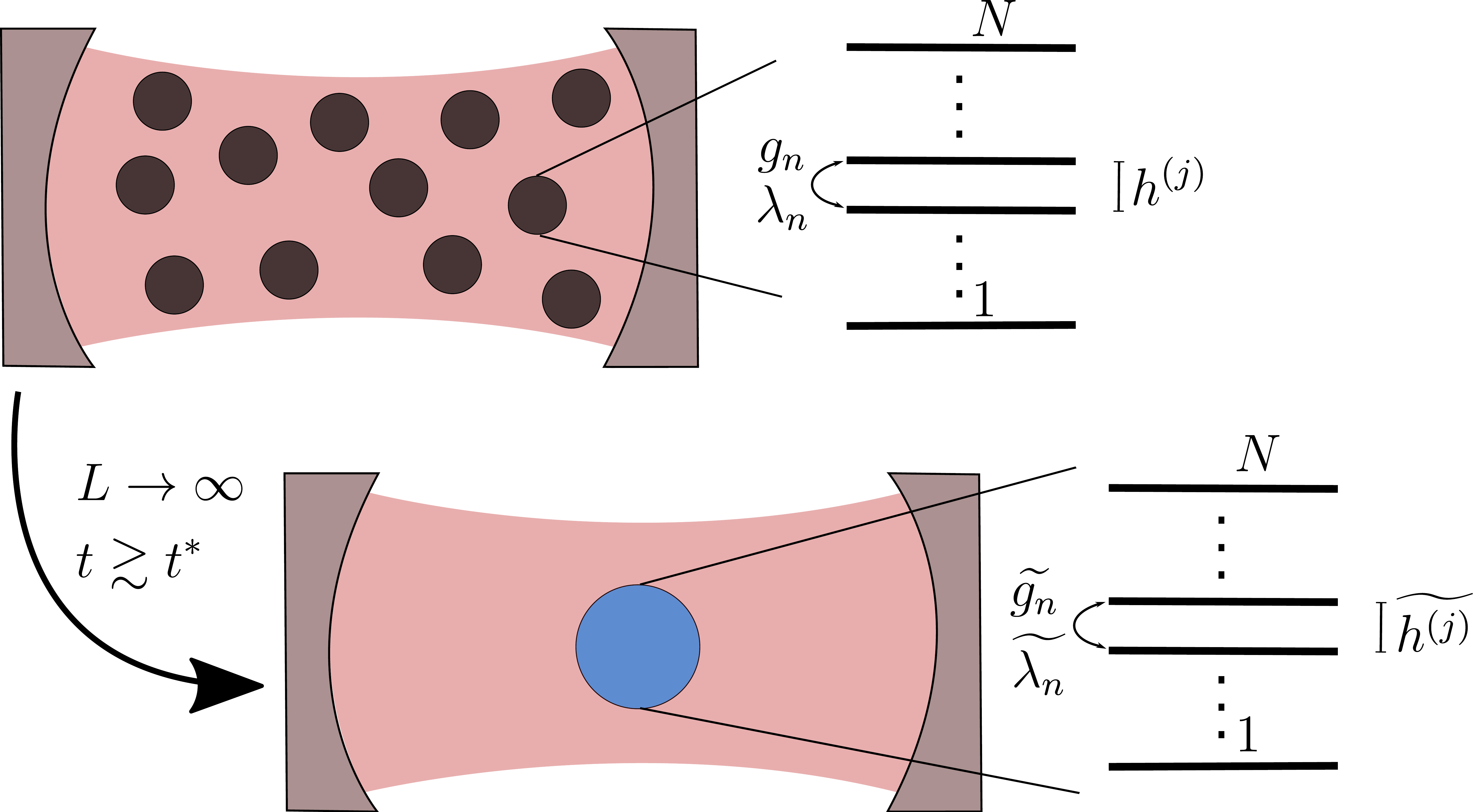}
\caption{Sketch of $\mathcal{N}_a$ atoms, each one hosting $N$ levels (panels on the side), distributed over $L$ sites (black dots), interacting via a common cavity field (red area). In the cartoon below we show the effective {$X$}-body system towards which the original many-body system is attracted in the long time $t \gtrsim t^*$. {We show a single-body effective model ($X=1$), since it is the one explicitly considered throughout our work.} We also show the internal structure of the single site both in the original many-body system and in the effective few-body description. \label{fig:redhop}}
\end{figure}

In Sec.~\ref{sec_semiclassical_limit} we argued that, in the $L\rightarrow \infty$ limit and for an initial state of the form given in Eq.~\eqref{eq_generic_state}, the dynamics of the cavity field and multilevel atoms are described by the equations of motion generated from a classical Hamiltonian composed of an extensive number (in the size $L$) of classical $SU(N)$ spins.
The dynamical reduction hypothesis conjectures that the dynamics of collective observables are effectively described by a classical Hamiltonian composed of a finite number, $X$, of effective $SU(N)$ systems~(cf.  Fig.~\ref{fig:redhop}); or in other words, the emergent collective dynamics can be effectively captured by a few-body macroscopic system.

Specifically, we conjecture that a fully-connected many-body system with $L$-sites, each with local degrees of freedom $\textbf{s}_j=\{s_{j,1},s_{j,2},...\}$, and classical Hamiltonian $H(\{\textbf{s}_j\}_{j=1}^L)$ will, after a sufficiently long time $t \gtrsim t^*$ and in the thermodynamic limit $L\to\infty$, possess an effective $X$-site effective model describing the collective dynamics.
The hypothesis supposes that the effective model will have $X$ finite, even when $L$ is in thermodynamics limit,  and that the effective local degrees of freedom $\{\widetilde{\textbf{s}}_j\}_{j=1}^X$  will be governed by a classical effective Hamiltonian $\widetilde{H}(\{\widetilde{s}_j\}_{j=1}^X)$. 
Hence, in order to predict dynamics of a collective observable $S(t)=f(\{\textbf{s}_j(t)\}_{j=1}^L)$, we will assume the existence of a   function $\tilde{f}$ of the effective degrees of freedom, which  will effectively reproduce  the dynamics of $S(t)$. Note that in general,  $\widetilde{f}$ is not necessarily of the same functional form of $f$. We can then compactly formulate the dynamical reduction hypothesis as
\begin{equation}
\label{eq_reduction_hypothesis}
\begin{split}
\lim_{L\to\infty} H(\{\textbf{s}_j\}_{j=1}^L) &\xrightarrow{t \gtrsim t^*} \widetilde{H}(\{\widetilde{\textbf{s}}_j\}_{j=1}^X),\\
\lim_{L\to\infty}S=f(\{\textbf{s}_j\}_{j=1}^L)  &\xrightarrow{t \gtrsim t^*} \widetilde{f}(\{\widetilde{\textbf{s}}_j\}_{j=1}^X).
\end{split}
\end{equation}
The effective Hamiltonian $\widetilde{H}$ is of the same functional form in the integrable case~\cite{PhysRevLett.96.097005,PhysRevA.91.033628,PhysRevB.99.054520}, while it is not generally expected to be so for non-integrable systems~\cite{PhysRevB.99.054520}. Importantly, we assume that the effective Hamiltonian obeys the same global symmetries as the many-body Hamiltonian.  

In the following we apply the dynamical reduction hypothesis~\eqref{eq_reduction_hypothesis} to craft various dynamical responses associated to the problem of synchronization in bosonic multivel cavity QED   summarized in Fig.~\ref{fig:cartoon}. We believe that our conjecture has universal flavor and it is applicable  to several other settings as we elaborate further in the concluding section.
\subsection{Classification of Dynamical Responses \label{sec_dynamical_phases}}
We construct a classification of dynamical phases by considering the different dynamics collective observables can display {in the many-body} system.
In the case of cavity QED, we consider the magnitude of the intra-level average coherence $|\Sigma_{n,m}(t)| = |\sum_{j=1}^L \Sigma_{n,m}^{(j)}(t)|/{L}$ with $n \neq m$. 
These observables can distinguish between cases when the atoms are synchronized {($|\Sigma_{n,m}(t)| \neq 0$)} or desynchronized {($|\Sigma_{n,m}(t)| = 0$)}, and in the case of synchronization we distinguish four dynamical responses.\\
\\
Desynchronized phase:
\begin{itemize}
\item Phase-I: in the long time limit {$|\Sigma_{n,m}(t)| \to 0$}, as a result of classical dephasing processes in the microscopic model due to  inhomogeneities in the local fields; 
\end{itemize}
Synchronized phases:
\begin{itemize}
\item Phase-II: $|\Sigma_{n,m}(t)|$ relaxes to a stationary {non-zero} value;
\item Phase-III: $|\Sigma_{n,m}(t)|$ displays self-generated Floquet dynamics (i.e. periodic oscillations) {characterized by a spectrum with well-resolved commensurate frequencies};
\item Phase-IV:  $|\Sigma_{n,m}(t)|$ displays aperiodic oscillations characterized by a spectrum with  well-resolved incommensurate frequencies;
\item Phase-IV${}^\star$: $|\Sigma_{n,m}(t)|$ displays chaotic oscillations exponentially sensitive to small changes in the initial conditions {and characterized by a spectrum with multiple broad peaks}.
\end{itemize}
While Phase-I and Phase-II are quite generic in the presence of inhomogeneous dephasing, Phase-III, Phase-IV and Phase-IV${}^\star$ are examples of   self-generated non-relaxing responses in absence of an  external drive.
As previously mentioned, the dynamical responses from Phase-I to Phase-IV were already observed in the integrable two-level case~\cite{PhysRevLett.93.160401,PhysRevLett.96.230403,PhysRevLett.126.173601,PhysRevA.91.033628}, while the chaotic Phase-IV${}^\star$ is accessible only in non-integrable systems~\cite{PhysRevB.104.104505}.

To predict and control when such phases occur we use the dynamical reduction hypothesis, and arguments based on symmetry and the Liouville-Arnold theorem.
The Liouville-Arnold theorem~\cite{goldstein2002classical} states that given a system with $M$ degrees of freedom and $Q$ conserved quantities, there exists a canonical transformation through `action-angle' variables, such that $Q$ `actions' are constant, and $Q$ `angles' evolves periodically at a frequency imposed by the value of the corresponding conserved quantity~\cite{Arnold1978,goldstein2002classical,Babelon2003}. Thus,
if $2Q \geq M$, the dynamics is solely along tori and the system is said to be classically integrable.
If instead $2Q< M$, there will be $(M-2Q)$ degrees of freedom  which evolve without any constraint and can in principle display chaotic behavior.
Notice that $Q\geq 1$ since the effective Hamiltonian always obeys time translation symmetry such that the effective energy is always a conserved quantity. 

To apply this theorem to describe the different phases with different effective models, we assume that an $X$ site effective model has in total $M$ DOFs.
Phase-I can be described by an effective model with $X=0$ sites, thus $M=0$ DOFs, since no effective degree of freedom  is necessary to capture a vanishing observable.
In the microscopic models, the synchronized phases generally occurs when the all-to-all coupling is large enough with respect to the inhomogeneities in the local fields, and it can be captured by an effective model with $X\geq 1$ sites, thus,  $M\geq 1$ DOFs, since we need at least one DOF for describing   nontrivial behavior. Combining the number of DOFs $M$, the number of symmetries $Q$, and the number of symmetries under which the specific observable is invariant, it is possible to predict the specific synchronized dynamical response. We show that, in $SU(N)$ systems, an effective single-site Hamiltonian ($X=1$) is already sufficient for observing all the dynamical responses from Phase-I up to Phase-IV${}^\star$.
This is in contrast to the $SU(2)$ integrable case in which an $X$-body effective Hamiltonian is necessary to capture Phase-$(X+1)$~\cite{PhysRevLett.93.160401,PhysRevLett.96.230403,PhysRevLett.126.173601,PhysRevA.91.033628,kelly2022resonant,PhysRevLett.126.173601}.

To apply this classification to multilevel cavity QED, we must identify the global symmetries present in such systems and the number of DOFs that could occur in the effective models.
We identify the global symmetries and number of DOFs in Sec.~\ref{sec_counting}, present a few  examples of effective models in Sec.~\ref{sec_eff_models} and give the predictions for the allowed dynamical responses for different $N$-level systems in Sec.~\ref{sec_QEDphases}.
\subsection{Counting DOFs and symmetries}\label{sec_counting}
Given a generic product state, as in Eq.~\eqref{eq_generic_state},
we conjecture an effective classical model composed of effective DOFs describing the matter and the cavity field separately. 
We assume that the effective cavity field is given by a bosonic amplitude $\widetilde{a}$ specified by two real numbers. As the detuning  from the atomic transitions increases, the contribution from such DOFs becomes suppressed, and consequently can be neglected in the far detuned limit~\cite{kelly2022resonant}. The effective matter's DOFs are either $SU(N)$ spins or bosonic amplitudes, depending on whether the collective observables $\Sigma_{n,m}(t)$ can be factorized or not.

If $\Sigma_{n,m}(t)$ cannot be factorized, the emergent effective classical model is composed of $X$ $SU(N)$-spins with elements $\widetilde{\Sigma}_{n,m}^{(k)}$ where $n,m \in [1,N]$ and $k\in [1, X]$.
Such an effective model has  $M = X \times N^2$ matter DOFs,
corresponding to the $N^2$ matrix elements for each effective spin $\widetilde{\Sigma}^{(k)}$. 
Since the effective degrees of freedom are $SU(N)$ spins, the number of independent parameters is reduced due to the Casimir charges  $\sum_{n=1}^N \widetilde{\Sigma}_{n,n}^{(k)}$ and $\sum_{n,m=1}^N \widetilde{\Sigma}_{n,m}^{(k)}\widetilde{\Sigma}_{m,n}^{(k)}$, which are the conservation of the number of bosons and length of the $SU(N)$ spin on each site $k$. As a consequence, the number of independent matter DOFs is $M=X \times (N^2-2)$.

Instead, if $\Sigma_{n,m}(t)$ can be factorized, then the effective model in the $SU(N)$ spins further simplifies and involves only $X \times N$ effective bosonic amplitudes $\widetilde{b}_{n,k}$ with $k\in [1, X]$ and $n\in[1,N]$. In this case, the number of matter DOFs is $M = X\times 2N$, being each bosonic amplitude specified by two real parameters. Assuming that the effective $SU(N)$ spins and effective bosons are related analogously to the microscopic ones via $\widetilde{\Sigma}_{n,m}^{k} = \widetilde{b}_{n,k}^*\widetilde{b}_{m,k}$, the two Casimir {charges} defined above are still conserved. In this case they are dependent one from the other and can be linked to the local $U(1)$ symmetry $\widetilde{b}_{n,k}\rightarrow  e^{i\phi_k}\widetilde{b}_{n,k}$ of the bilinears $\widetilde{b}_{n,k}^* \widetilde{b}_{m,k}$. Since the number of bosons is conserved, the corresponding conjugate variable, the sum of the phases of the bosonic amplitudes, is irrelevant and the number of nontrivial matter DOFs is $M=X\times (2N-2)$. 

Once the effective DOFs are identified, we can construct the effective Hamiltonian which governs their dynamics imposing the same symmetries of the many-body Hamiltonian in Eq.~\eqref{eq_H_SUn_wphoton} and Eq.~\eqref{eq_H_sun_adiabatic} in the classical limit.
The first symmetry is time translation invariance, which implies the conservation of the energy, while the second is a global $U(1)$ symmetry present solely in absence of co-rotating processes. 
Specifically, for $\lambda_n=0$ the Hamiltonian in Eq.~\eqref{eq_H_SUn_wphoton} is invariant under $(\Sigma_{n,n+1},a) \to (e^{i\theta}\Sigma_{n,n+1},e^{i\theta}a)$ and thus conserves the number of total excitations, which in the two-level case is $[(\Sigma_{2,2}-\Sigma_{1,1})/2+|a|^2]$ while in the generic multilevel case is a linear combination of $\{\Sigma_{n,m}\}$ and $|a|^2$~\cite{campos2021generalization,campos2022generalization}. Analogously, for $\nu_{m,n},\zeta_{n,m} = 0$ the atoms-only model in Eq.~\eqref{eq_H_sun_adiabatic} is invariant under $\Sigma_{n,n+1} \to e^{i\theta}\Sigma_{n,n+1}$, which leads to the conservation of the number of atomic excitations (e.g. $(\Sigma_{2,2}-\Sigma_{1,1})$ in the two-level case).

Combining the effective DOFs and symmetries, we can now propose a possible set of effective models and predict the dynamical responses of collective observables via arguments based on symmetry and the Liouvile-Arnold theorem.

\begin{table}[]
\begin{tabular}{l|l|l|l|l}
                                       & {$Q$} & $N=2$           & $N=3$           & $N \geq 4$           \\ \hline
$\widetilde{g},\widetilde{\lambda} \neq 0$ and $\widetilde{\omega}_0$ finite   & {1} & IV${}^{\star}$  & IV${}^{\star}$  & IV${}^{\star}$\\
$\widetilde{g},\widetilde{\lambda} \neq 0$ and $\widetilde{\omega}_0\rightarrow \infty$ & {1} &III           & IV${}^{\star}$  & IV${}^{\star}$\\
$\widetilde{\lambda} =0$ and $\widetilde{\omega}_0$ finite       & {2}         & III           & IV${}^{\star}$  & IV${}^{\star}$ \\
$\widetilde{\lambda} =0$ and $\widetilde{\omega}_0\rightarrow \infty$ & {2} & II            & III or IV${}^{\star}$   & IV${}^{\star}$ 
\end{tabular}

\caption{Summary of the dynamical responses of the magnitude of the intra-level average phase coherence captured by the effective Hamiltonians in Eq.~\eqref{eq_H_noadiab_W0} and Eq.~\eqref{eq_H_W0}. {The number of matter DOFs is either $(2N-2)$ or $(N^2-2)$ depending on whether $\widetilde{\Sigma}$ can be factorized or not, respectively (cf. Sec.~\ref{sec_counting}). 
If the cavity field detuning $\widetilde{\omega}_0$ is finite, we need two additional DOFs to describe the modulus and phase of the actively participating cavity field.}
The presence of a $U(1)$ symmetry increases the number of conserved quantities $Q$ by $1$. {For the $N=3$ spin exchange model (last row) the system can display from Phase-I to either Phase-III or Phase-IV${}^\star$ depending on whether $\widetilde{\Sigma}$ can be factorized or not, respectively.}
In all cases, all the responses with `less order' than the one reported, could be in principle   accessed tailoring the initial state and the parameters of the Hamiltonian. The same table holds in the case the Hamiltonian is spatially homogeneous, since the effective models are trivially equal to the microscopic ones (cf. Sec.~\ref{sec_W0}). \label{table}}
\end{table}
\subsection{Effective Models}\label{sec_eff_models}
In order to make concrete the above picture, here we present   a set of possible effective models for multilevel cavity QED systems described by Eq.~\eqref{eq_H_SUn_wphoton}.
As mentioned above,  an exact derivation is not available  in the generic multilevel case (see  Refs.~\cite{PhysRevLett.96.097005,PhysRevA.91.033628,PhysRevB.99.054520} where the effective few-body Hamiltonian can be derived from the Richardson-Gaudin integrability of the SU(2) case).
Nonetheless, considering the initial state to be a generic product state (cf.  Eq.~\eqref{eq_generic_state}), the effective DOFs are $SU(N)$ spins, and the simplest effective theory is given by the microscopic Hamiltonian in Eq.~\eqref{eq_H_SUn_wphoton} with $L=1$  (thus $X=1$ effective sites)
\begin{equation}
\label{eq_H_noadiab_W0}
\begin{split}
    &\widetilde{H}(\widetilde{\Sigma}_{n,m},\widetilde{a})= \widetilde{\omega}_0 \widetilde{a}^* \widetilde{a}+\sum_{n=1}^N \widetilde{h}_n\widetilde{\Sigma}_{n,n}+ \\
&+\sum_{n=1}^{N-1} \Big[ \widetilde{g}_{n} \left(\widetilde{\Sigma}_{n+1,n}\widetilde{a} + h.c.\right)+ \widetilde{\lambda}_{n} \left(\widetilde{\Sigma}_{n+1,n}\widetilde{a}^* + h.c.\right)\Big].
\end{split}
\end{equation}
Analogously, in the far-detuned cavity mode limit described by the Hamiltonian in Eq.~\eqref{eq_H_sun_adiabatic}, we propose the effective Hamiltonian
\begin{equation}
\label{eq_H_W0}
\begin{split}
    &\widetilde{H}_{e}(\widetilde{\Sigma}_{n,m}) =\sum_{n=1}^N \widetilde{h}_n\widetilde{\Sigma}_{n,n}+\\
 &-\sum_{m,n=1}^{N-1}\Big[\widetilde{\chi}_{n,m}\widetilde{\Sigma}_{n+1,n} \widetilde{\Sigma}_{m,m+1} +\widetilde{\zeta}_{n,m} \widetilde{\Sigma}_{n,n+1} \widetilde{\Sigma}_{m+1,m}+\\
 &+\widetilde{\nu}_{n,m} \widetilde{\Sigma}_{n+1,n} \widetilde{\Sigma}_{m+1,m} + \widetilde{\nu}_{m,n}\widetilde{\Sigma}_{n,n+1} \widetilde{\Sigma}_{m,m+1}\Big].
\end{split}
\end{equation}
Additionally, if the collective observables $\widetilde{\Sigma}_{n,m}$ can be factorized, we   conjecture effective models for the boson DOFs of the form
\begin{eqnarray}
    \label{eq_reduced_bosonic_model_with_photon}
    \widetilde{H}(\widetilde{b}_{n},\widetilde{a})&=&\widetilde{H}(\widetilde{\Sigma}_{n,m}=\widetilde{b}_{n}^* \widetilde{b}_m,\widetilde{a}) \\ 
    \label{eq_reduced_bosonic_model}
    \widetilde{H}_{e}(\widetilde{b}_{n})&=&\widetilde{H}_{e}(\widetilde{\Sigma}_{n,m}=\widetilde{b}_{n}^* \widetilde{b}_m) 
\end{eqnarray}
where we conjecture that the effective one-body reduced density matrix factorizes as $\widetilde{\Sigma}_{n,m}=\widetilde{b}_n^* \widetilde{b}_m$.
These effective models are trivially exact when the Hamiltonians in Eq.~\eqref{eq_H_SUn_wphoton} and Eq.~\eqref{eq_H_sun_adiabatic} are spatially homogeneous for $h_n^{(j)}=h_n$ at $W=0$. Indeed, at $W=0$ the many-body Hamiltonians trivially reduces to a few-body one due to the permutation symmetry under swapping of any pair of sites. Despite their apparent simplicity, the effective models here introduced allow us to obtain quantitatively  the whole set of dynamical responses described in Sec.~\ref{sec_dynamical_phases}. Furthermore,  we show in Sec.~\ref{sec_Wfinite} that these models describe correctly the dynamics of collective observables also at moderate inhomogeneity, with a quantitative matching in the case of $N=3$  spin-exchange interactions.

\subsection{Classification for multilevel cavity QED}\label{sec_QEDphases}
We are now in the position to discuss the possible dynamical phases for the $X=1$ effective models introduced in Sec.~\ref{sec_eff_models} for different number of levels $N$.
As already anticipated, we consider as collective variable the magnitude of the intra-level average coherences, which in the effective models are given by $|\widetilde{\Sigma}_{n,m}(t)|$ with $n\neq m$.
The results of this section are summarized in Table~\ref{table}.

For a generic multilevel atom with $N\geq 4$ levels, the number of DOFs $M$ is always larger than the $2Q\leq 4$ symmetries identified.
Thus, generically, the effective model can show aperiodic oscillations~(Phase-IV) and may even display chaotic behavior~(Phase-IV${}^\star$).

\subsubsection{$N=2$ level atoms}
The   case of $N=2$ levels has been well studied~\cite{kirton2019introduction,Bogoliubov1996,GLICK1965211} and in this section we discuss how  our approach reproduces known results.
The number of matter DOFs is $M=2$, either considering an effective bosonic model, or with $SU(2)$ spins. {Thus, it is not possible to access different dynamical responses upon introducing quantum correlations in the mean-field limits considered}.  
If the cavity field actively participates to dynamics, we need to keep track of two additional DOFs given by the real and imaginary part of its amplitude.

Let us consider the two level system with a photon actively participating in  dynamics. We can identify the two regimes corresponding to either the generalized Dicke model ($\widetilde{\lambda},\widetilde{g}\neq 0$) or the Tavis-Cummings model ($\widetilde{\lambda}=0$). {Both models have $M=4$ DOFs, but a different number of conserved quantities.}
The generalized Dicke model conserves only the energy beyond the total spin (which we already taken into account), opening  the option of chaos (Phase-IV${}^\star$), as it has been seen for instance in Refs.~\cite{emary2003chaos,BastarracheaMagnani2015,PhysRevLett.122.024101,pilatowsky2020positive,PhysRevE.94.022209,alavirad2019scrambling,PhysRevE.93.022215,lerma2019dynamical,BastarracheaMagnani2017}. 
Instead, the Tavis-Cummings model has one additional conserved charge (total number of excitations), is therefore integrable and in fact it shows regular dynamics~\cite{Bogoliubov1996,Barmettler2013,PhysRevA.2.336,PhysRevA.79.053825}.
Under change to action-angle variables, the dynamics are seen as the evolution on a 3-tori, with 3 independent frequencies. Thus a general observable might show Phase-IV oscillations.
Nevertheless, the magnitude of the mean coherence $|\widetilde{\Sigma}_{1,2}(t)|$ 
{only shows periodic oscillations~(Phase-III) since it is} invariant under two of the symmetries, {specifically the $U(1)$ symmetries $(\widetilde{\Sigma}_{1,2},\widetilde{a})\to(e^{i\theta}\widetilde{\Sigma}_{1,2},e^{i\theta}\widetilde{a})$ and $\widetilde{b}_{n} \to e^{i\theta}\widetilde{b}_n$, with $n=\{1,2\}$, linked to the conservation of the total number of excitations and spin, respectively.}

In the limit where the cavity mode is far detuned from the atomic transitions, the Tavis-Cummings model becomes a simple spin-exchange model with $M=2Q=2$ (the conservation of energy and of the total number of excitations are dependent).
Since this model can only have two independent frequencies corresponding to the precession of the $U(1)$ angle variables, the observable $|\widetilde{\Sigma}_{1,2}(t)|$ is constant yielding Phase-II.
If the additional $U(1)$ symmetry is broken, one recovers the Lipkin-Meshkov-Glick (LMG)  model, and the observable $|\widetilde{\Sigma}_{1,2}(t)|$ can again oscillate with a single frequency and display Phase-III, as it is generically observed~\cite{scully1999quantum,PhysRevB.74.144423,PhysRevA.100.032117,PhysRevA.102.052210,Sciolla2011,PhysRevLett.121.240403}.

\subsubsection{$N=3$ level atoms}
In the three-level case, $M$ depends on  whether the model reduces to a bosonic model or to a $SU(3)$ spin system. In the bosonic case, the atomic sector is described by $M=2N-2=4$ real DOFs. In the generic $SU(3)$ case, the matter is described by $M=N^2-2=7$ DOFs.
If the photon is an active DOFs, its additional DOFs leads to $M\geq 6$ in either the bosonic or spin model, and since $Q \leq 2$ for any set of parameters, the dynamics can enter the chaotic Phase-IV${}^\star$.  

The case of the spin-exchange model, corresponding to $\widetilde{\zeta}_{n,m}=\widetilde{\nu}_{n,m}=0$ in Eq.~\eqref{eq_H_W0}, is perhaps the most interesting, since depending  on whether the model reduces to a bosonic model or a $SU(3)$ spin model, the dynamics can be either in Phase-III or Phase-IV (possibly IV${}^\star$).
Indeed, the three level bosonic model $\widetilde{H}_{e}(\widetilde{b})$ has $M=4$ DOFs and $Q=2$ conserved charges, corresponding to the total energy and `number of excitations' $(\widetilde{\Sigma}_{3,3}-\widetilde{\Sigma}_{1,1})$ 
, and it is therefore integrable.
Again, two of the frequencies are absorbed in the invariance of $|\widetilde{\Sigma}_{n,m}|$ under the two $U(1)$ symmetries, and thus all oscillations must be periodic, yielding Phase-III.
If instead the initial state has $\widetilde{\Sigma}_{nm}\neq\widetilde{b}_n^*\widetilde{b}_m$, we must consider the $SU(3)$ spin model $\widetilde{H}_{e}(\widetilde{\Sigma}_{n,m})$, and the extra number of degrees of freedom leads $M>2Q=4$ allowing the dynamics to be either Phase-IV or Phase-IV${}^\star$.
If $\widetilde{\zeta}_{n,m},\widetilde{\nu}_{n,m} \neq 0$ the system loses a $U(1)$ symmetry, associated to the conservation of $(\widetilde{\Sigma}_{3,3}-\widetilde{\Sigma}_{1,1})$, and consequently can display chaotic behavior for both bosons or $SU(3)$ spins.

\section{Homogeneous systems \label{sec_W0}}
In this section, we consider the homogeneous case $(W=0)$ of Eq.~\eqref{eq_H_sun_adiabatic}, where the dynamical reduction hypothesis is true due to the permutation symmetry, and we test the predictions of Table~\ref{table}. 
In Sec.~\ref{sec_homogeneous_bosonic} we consider permutation invariant coherent states, discussing the role of the interactions. In Sec.~\ref{sec_chaos_induced_fluctuactions} we discuss the consequences of classical and quantum correlations in the initial state focusing on the $N=3$ levels spin-exchange Hamiltonian, where we observe the onset of a chaotic phase.\\

For $W=0$ both the Hamiltonians in Eqs.~\eqref{eq_H_SUn_wphoton},~\eqref{eq_H_sun_adiabatic} are permutationally invariant under swapping of any pair of sites, and they can be written as a function of the collective operators in Eq.~\eqref{eq_collective_operators}. Thus, we can immediately achieve the thermodynamic limit $L\to\infty$ considering a single large $SU(N)$ spin. As a consequence, in the mean field limit we exactly obtain the classical Hamiltonians $\widetilde{H}(\widetilde{\Sigma}_{n,m},\widetilde{a})$ or $\widetilde{H}_{e}(\widetilde{\Sigma}_{n,m})$, depending on whether the cavity field is an active DOF or not, respectively. 
Here, unlike in the general case, the effective DOFs trivially relate to the  {original} collective DOFs being $\Sigma_{n,m}=\widetilde{f}(\widetilde{\Sigma}_{n,m})=\widetilde{\Sigma}_{n,m}$, and the parameters of the effective Hamiltonians are equal to the {original} ones (e.g. $\widetilde{\omega}_0 = \omega_0$). 
As discussed in Sec.~\ref{sec_semiclassical_limit}, the choice of reducing  the model to spin degrees of freedom, i.e. $\widetilde{H}_{e}(\widetilde{\Sigma}_{n,m})$, or   to bosons, $\widetilde{H}_{e}(\widetilde{b}_n)$, depends only on the purity of the effective one-body reduced density matrix $\widetilde{\Sigma}$, i.e. whether    its elements $\widetilde{\Sigma}_{n,m}$ can be factorized as the product of bosonic operators. In the next sections we investigate both scenarios and confirm the prediction summarized in Table~\ref{table}.

\subsection{Homogeneous coherent states \label{sec_homogeneous_bosonic}}
We set as initial state a permutationally invariant (in space) coherent state with equal average occupation on each level ($\gamma_{n,j}=\gamma$ in Eq.~\eqref{eq_bosonic_coherent_state}). Since the state is homogeneous in space, the average one-body reduced density matrix is pure, and can be factorized in the bosonic amplitudes as $\Sigma_{m,n} = b_{m}^* b_n$. Therefore, the effective model describing collective observables is either $\widetilde{H}(b_n , a)$ or $\widetilde{H}_{e}(b_n)$ 
(notice that we interchanged the effective bosonic amplitudes with the microscopic ones being equal in this case). We show results only in the case where the photon is not an active DOF. Generally, we expect that an active photon leads to a change of dynamical response from Phase-$Y$, displayed in its absence, to Phase-$(Y+1)$, due to the additional DOFs~\cite{kelly2022resonant}. Nonetheless, this effect is suppressed as the detuning of the cavity field frequency with respect to the atomic transitions increases, making our results approximately valid also for large but finite detunings.

\begin{figure}[t!]
\centering
\includegraphics[width=\linewidth]{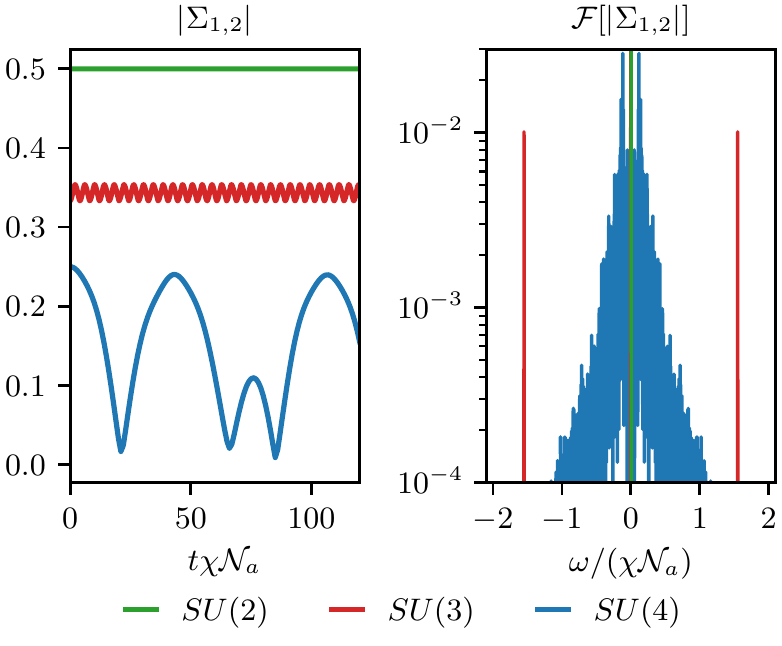}
\caption{Left plot: Dynamics of the magnitude of the average phase coherence 
$|\Sigma_{1,2}(t)|$ 
in the homogeneous limit ($W=0$) for the {$N$-levels} spin-exchange model (Eq.~\eqref{eq_H_sun_adiabatic} with $\nu_{n,m}=\zeta_{n,m}=0$) with $N=\{2,3,4\}$. The initial state is a permutation invariant (in space) coherent state with same average occupation on each level $n \in [1,N]$. The couplings $\chi_{n,m}$ are chosen such that we have genuine $SU(N)$ {spins}. 
For $N=2$, $|\Sigma_{1,2}(t)|$ is constant (Phase-II); for $N=3$, $|\Sigma_{1,2}(t)|$ displays periodic oscillations (Phase-III); for $N=4$, there are not enough conserved quantities to constrain  the space of accessible states and therefore $|\Sigma_{1,2}(t)|$ displays oscillations exponentially sensitive to small changes in   initial conditions (Phase-IV${}^\star$). Right plot: {magnitude of the} Fourier spectrum of $|\Sigma_{1,2}(t)|$. 
For $N=2$ the only nonzero component is at $\omega=0$, being $|\Sigma_{1,2}(t)|$ a constant; for $N=3$ the spectrum has two well-resolved peaks; for $N=4$ there are multiple broad peaks. Due to the permutation invariance under swapping of any pair of sites, the thermodynamic limit can be achieved by simulating a single site ($L=1$) in the microscopic model in Eq.~\eqref{eq_H_sun_adiabatic}, or equivalently the effective model of Eq.~\eqref{eq_reduced_bosonic_model}.}
\label{fig_dynamics_homogeneous_initial_state_W0}
\end{figure}

In Fig.~\ref{fig_dynamics_homogeneous_initial_state_W0} we show the dynamics of the magnitude of the phase coherence $|\Sigma_{1,2}|$ in the spin-exchange model (Eq.~\eqref{eq_H_sun_adiabatic} with $\nu_{n,m}=\zeta_{n,m}=0$) in the homogeneous limit ($W=0$) for different number of atomic levels $N \in \{2,3,4\}$. As predicted in Table~\ref{table}, upon changing   the number of levels, the system can display markedly different dynamical responses: for $N=2$, $|\Sigma_{1,2}|$ displays Phase-II; for $N=3$, $|\Sigma_{1,2}|$ displays Phase-III; for $N=4$, $|\Sigma_{1,2}|$ displays chaotic behavior (Phase-IV${}^\star$).  
We also observe the onset of aperiodic oscillations (Phase-IV) in the {$N=4$ } case for different sets of parameters and initial states. 
This is a simple signature of the importance of considering multilevel atoms, although the model has all-to-all interactions.

\subsection{{Chaos induced by quantum correlations} \label{sec_chaos_induced_fluctuactions}}
In this section we discuss the impact of quantum {correlations} in the initial state in the spin-exchange model (Eq.~\eqref{eq_H_sun_adiabatic} with $\nu_{n,m}=\zeta_{n,m}=0$). 
We show that the subsequent dynamics is susceptible to quantum {correlations}, with particularly striking effects in the $SU(3)$ case, where we can craft a specific dynamical response {by manipulating the initial state}, from Phase-III up to Phase-IV${}^\star$ (chaos).\\ 

\begin{figure}[t!]
\centering
 \includegraphics[width=\linewidth]{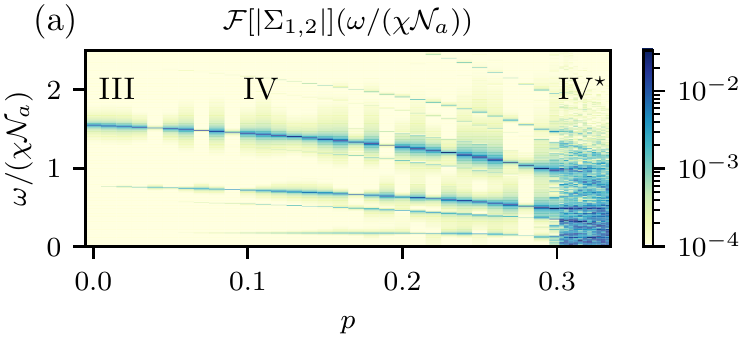}\\
	\includegraphics[width=0.335\linewidth]{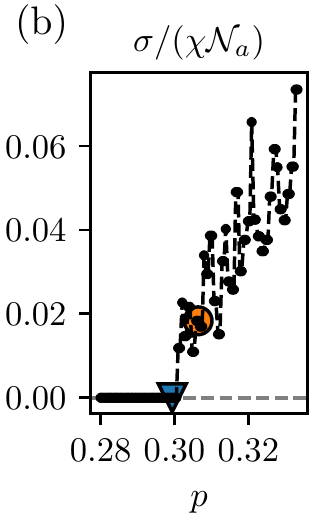}
        \includegraphics[width=0.63\linewidth]{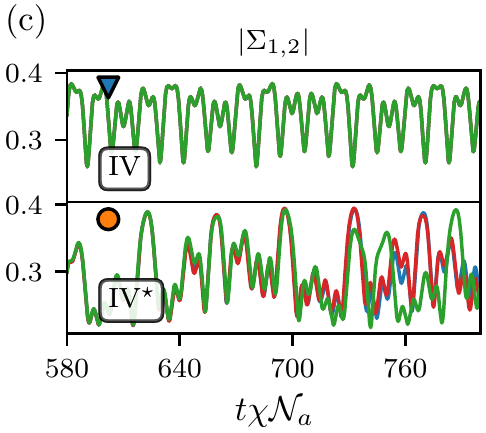}
\caption{
{Dynamical response in the $N=3$ levels spin-exchange model starting from a multimode Schr\"{o}dinger cat state (cf. Eq.~\eqref{eq_parametrization_quantum_chaotic_state}) as a function of initial quantum correlations parameterized via $p$. We set $\mathcal{N}_a\to\infty$, $g_1/g_2 \approx 2$ and $W=0$. (a) magnitude of the Fourier spectrum of the magnitude of the average coherence $|\Sigma_{1,2}|$ (the other $|\Sigma_{n,m}|$ behaves similarly) as a function of $p$, which displays a crossover from few commensurate peaks (Phase-III) to a regime with multiple incommensurate ones (Phase-IV), and eventually signals the onset of a chaotic phase (Phase-IV${}^\star$). (b) Maximum Lyapunov exponent $\sigma/(\chi\mathcal{N}_a)$ as a function of $p$, which enables to locate the transition from a regular regime to a chaotic one at $p^\star \approx 0.3$. (c) Dynamics of $|\Sigma_{1,2}(t)|$ starting from three nearly sampled initial states at two different values of $p\approx\{0.299,0.306\}$ (marked in (b)), showing exponential sensitivity to changes in  initial conditions in Phase-IV${}^\star$.}}
\label{fig_classical_fluctuactions_chaos}
\end{figure}

{We set on each site the same multimode Schr\"{o}dinger cat state (cf. Eq.~\eqref{eq_generic_entangled_state}). As discussed in Sec.~\ref{sec_initial_states}, when $|\langle \widetilde{\gamma}^{(1)}|\widetilde{\gamma}^{(2)}\rangle| < 1$, the one-body reduced density matrix is mixed, the phase-coherences do not factorize ($\Sigma_{n,m} \neq b_n^* b_m$), and 
we have to keep track of all the bilinears $\Sigma_{n,m}$.
Thus, the effective model passes from the one in the bosonic DOFs defined in Eq.~\eqref{eq_reduced_bosonic_model} to the one in the $SU(N)$ spins defined in Eq.~\eqref{eq_H_W0}. Due to quantum correlations, the number of effective DOFs passes from $2N$ to $N^2$ and the constraint imposed by the conserved quantities no longer ensures classical integrability for $N>2$. This is particularly striking in the $SU(3)$ case, where quantum correlations in the initial state can lead to a transition from a regular regime to a chaotic one. For this reason, we focus on the $SU(3)$ case in the following. Specifically, we consider as initial state a family of multimode Schr\"{o}dinger cat state (cf.  Eq.~\eqref{eq_generic_entangled_state}) parameterized via a parameter $p\in[0,1/3]$ as
\begin{equation}
\label{eq_parametrization_quantum_chaotic_state}
\begin{split}
\boldsymbol{\gamma}^{(1)} &= \sqrt{\frac{\mathcal{N}_a}{L}} \cdot (\sqrt{1/3+p},\sqrt{1/3},\sqrt{1/3-p}),\\
\boldsymbol{\gamma}^{(2)} &= \sqrt{\frac{\mathcal{N}_a}{L}} \cdot (\sqrt{1/3},\sqrt{1/3-p},\sqrt{1/3+p}).
\end{split}
\end{equation}
The overlap $|\langle \widetilde{\gamma}^{(1)}|\widetilde{\gamma}^{(2)}\rangle|$ is exponentially suppressed both in $p$ and $\mathcal{N}_a/L$, so that $\langle \widetilde{\gamma}^{(1)}|\widetilde{\gamma}^{(2)}\rangle=0$ for any $p >0$ in the limit $\mathcal{N}_a/L \to \infty$.
We quantify quantum correlations by the connected two-point functions $(\Sigma_{n,m}-b_n^* b_m)$. Given the state in Eq.~\eqref{eq_parametrization_quantum_chaotic_state}, the connected two-point functions are null at $p=0$ and increase polynomially with $p$. As a consequence, the number of effective DOFs $M$ needed is expected to increase with $p$. Based on our classification, we thus expect a change of the collective dynamical response displayed. 
This is manifest looking at the Fourier spectrum of $|\Sigma_{1,2}|$ (cf. Fig.~\ref{fig_classical_fluctuactions_chaos}(a)), where as $p$ increases we observe a crossover from a regime with few commensurate peaks (Phase-III) to a regime with multiple incommensurate one (Phase-IV), analogous to period doubling phenomena, and eventually the onset of chaos (Phase-IV${}^\star$) for $p \gtrsim p^\star$. The value $p^\star$ generally depends on the parameters of the Hamiltonian. We locate $p^\star$ computing the maximum Lyapunov exponent $\sigma$, which is the largest exponential rate at which nearby trajectories diverge and it is finite and positive in chaotic system and zero for regular Hamiltonian dynamics~\cite{PhysRevA.96.023624,gaspard2005chaos}. We find $p^\star \approx 0.3$ for $g_1/g_2 \approx 2$ (cf Fig.~\ref{fig_classical_fluctuactions_chaos}(b)).
We refer to Appendix~\ref{appendix_computation_lyapunov} for the details about the calculation of the Lyapunov exponent and $p^\star$.}

We highlight that the interactions between $SU(3)$ spins are an essential ingredient for observing chaos. Indeed, for $g_1=g_2$ (and thus $\chi_{1,1}=\chi_{2,2}=\chi_{1,2}$) dynamics take place in a $SU(2)$ subgroup of $SU(3)$, thus the number of DOFs reduces and there cannot be chaos as a consequence of the Arnold-Liouville theorem. While deep in the $SU(3)$ regime we have chaos for any $p \gtrsim p^\star$, instead near the $SU(2)$ limits we observe regions in $p$ of chaotic behavior embedded in regular ones (specifically Phase-IV).
In Appendix~\ref{appendix_dynamical_response_chaos_SU3} we provide details on the Lyapunov exponent as a function of $p$ and the ratio $g_1/g_2$, passing from  the $SU(2)$ ($g_1=g_2$) to the $SU(3)$ spin case ($g_1 \neq g_2 \neq 0$).

In absence of interference effects ($\langle \widetilde{\gamma}^{(1)}|\widetilde{\gamma}^{(2)}\rangle = 0$), as it is for any $p>0$ in the limit $\mathcal{N}_a/L\to\infty$ considered, the equations of motion of the collective observables $\Sigma_{n,m}$ are the same starting either from the Schr\"{o}dinger cat state in Eq.~\eqref{eq_generic_entangled_state}, or from a state with half sites in the state $|\widetilde{\gamma}^{(1)}\rangle$ and the other half in the state $|\widetilde{\gamma}^{(2)}\rangle$.
In this context, $p$ effectively controls the `sharpness' of a `kink' in the initial spatial configuration of the $SU(3)$ spins, in analogy with domain walls in the $SU(2)$ case~\cite{PhysRevLett.126.173601}. 
The primary difference is that in Ref.~\cite{PhysRevLett.126.173601} it is only possible to generate Phase-III by considering an inhomogenous configuration of the local fields.
Specifically, they consider a configuration such that the local fields are positive in half the sites and negative in the other half, and initializing the $z$-component of the spins along their corresponding local field, which is equivalent to a spatial `kink'. In analogy, we can notice that embedding a Schr\"{o}dinger cat state is similar to the insertion of an internal `quantum kink': the word  `quantum' highlights the presence of multi-particles entangled states, while `kink' refers to the   phase-space representation of the state, which would be given by two coherent states pointing in opposite directions, but now in the internal Hilbert space of the atom. 

The sharp feature in the Lyapunov exponent as a function of $p$ in Fig.~\ref{fig_classical_fluctuactions_chaos}{(b)} looks similar to a first-order phase transition.  A field theory investigation of this phenomenon is ongoing and it represents  a natural and fruitful direction of outreach of our results.
\subsection{Chaotic seeds in   initial states \label{sec_chaotic_seed}}
We now explore  the option to induce a  chaotic phase   by initializing     a fraction of the sites in a Schr\"{o}dinger cat state, while keeping the  other sites in a coherent state.
We consider $|\psi_\text{cat}\rangle \sim (|\widetilde{\gamma}^{(1)}\rangle + |\widetilde{\gamma}^{(2)}\rangle)$ as defined in Eq.~\eqref{eq_parametrization_quantum_chaotic_state} with $p\in[0,1/3]$.
We initialize a fraction $F\in [0,1]$ of sites in $|\psi_\text{cat}\rangle$ such that the initial state is
\begin{equation}
\label{eq_fraction_chaotic_state}
|\Psi\rangle = \otimes_{j=1}^{\lfloor FL\rfloor}|\psi_\text{cat}\rangle \otimes_{j=\lfloor FL\rfloor + 1}^L |\widetilde{\gamma}^{(1)}\rangle,
\end{equation}
where $\lfloor x\rfloor$ returns the least integer greater than or equal to $x$. 
The region initialized in $|\psi_\text{cat}\rangle$ could favor Phase-IV${}^\star$, while the region initialized in a coherent state would favor a regular dynamical response (Phase-III). 
We observe that the chaotic region proliferates  and drives the whole system into the chaotic Phase-IV${}^\star$ for $F \gtrsim F^\star$, where $F^\star$ depends on the details of the initial   state   and parameters of the Hamiltonian. For the state in Eq.~\eqref{eq_fraction_chaotic_state}, $F^\star \approx 0.5$ at $p\approx 1/3$. In  Appendix~\ref{appendix_finite_fraction_chaos} we offer a more detailed analysis.

\section{Effects of inhomogeneous fields \label{sec_Wfinite}}
Upon introducing inhomogeneous local fields ($W>0$) the permutation symmetry is broken and the Hamiltonians in Eq.~\eqref{eq_H_SUn_wphoton} and Eq.~\eqref{eq_H_sun_adiabatic} cannot be straightforwardly written as a function of collective DOFs. 
Regardless, the dynamical responses observed in the homogeneous case ($W=0$) are generally robust against finite inhomogeneity ($W>0$), and we provide numerical evidence that the simple effective Hamiltonians defined in Sec.~\ref{sec_eff_models}   describe quantitatively the many-body collective dynamics of the full model in a regime of moderate inhomogeneity $W$. 
In particular, we focus on the spin-exchange Hamiltonian in Eq.~\eqref{eq_H_sun_adiabatic} for $\nu_{n,m}=\zeta_{n,m}=0$, 
\begin{equation}
\label{eq_SUN_W_neq_0}
\begin{split}
H &=\sum_{j=1}^L\sum_{n=1}^N h_n^{(j)} \Sigma_{n,n}^{(j)}-\sum_{m,n=1}^{N-1}\chi_{n,m} \Sigma_{n+1,n} \Sigma_{m,m+1}.
\end{split}
\end{equation}

\begin{figure}[t!]
\centering
\includegraphics[width=\linewidth]{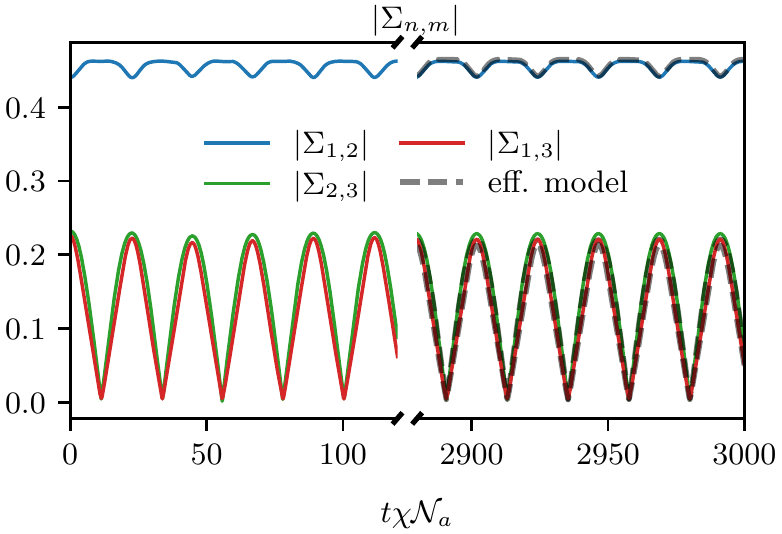}
\caption{Dynamics of the magnitude of the average phase coherence
$|\Sigma_{m,n}(t)|$ 
in the {$N=3$ levels} spin-exchange model at fixed $g_2/g_1 \approx 10^{-2}$ 
and $W/(\chi\mathcal{N}_a)=0.1$. The initial state is a permutationally invariant (in space) coherent state. The continuous lines are obtained simulating the full many-body dynamics with $L=10^4$ sites. The dashed black lines are obtained simulating the effective model in Eq.~\eqref{eq_H_W0} with parameters numerically obtained by the optimization of the cost function in Eq.~\eqref{eq_cost_function}.}
\label{fig_fit_phaseIII_in_SU3}
\end{figure}

To demonstrate the validity of the effective Hamiltonian, we numerically identify the parameters of the simple \textit{ansatz} in Eq.~\eqref{eq_H_W0}  that reproduce the dynamics of   collective observables. Our procedure   can be summarized as follows: 
\begin{enumerate}[(i)]
\item we compute the time evolution of the collective observables $\Sigma_{n,m}(t)$ from the full many-body dynamics obtained via the Hamiltonian in Eq.~\eqref{eq_SUN_W_neq_0}; 
\item  we set the initial conditions  $\{\widetilde{\Sigma}_{n,m}(t=0)\}$ and give a numerical `seed' to the parameters $\{\widetilde{h}_n,\widetilde{\chi}_{n,m},\widetilde{\zeta}_{n,m},\widetilde{\nu}_{n,m}\}$ in the effective model in~\eqref{eq_H_W0};
\item we compute the time evolution of the collective observables using the effective model in Eq.~\eqref{eq_H_W0};
\item we vary the initial conditions and effective Hamiltonian parameters to minimize the average norm-1 distance between $\Sigma(t)$ computed in (i) and $\widetilde{\Sigma}(t)$ computed using the effective model in (iii), i.e. we set as cost function
\begin{equation}
\label{eq_cost_function}
{\epsilon_1=}\frac{1}{T}\int_0^T\sum_{n,m=1}^N\left|\widetilde{\Sigma}_{n,m}(t) - \Sigma_{n,m}(t)\right|\:dt.
\end{equation}
\end{enumerate}
In Appendix~\ref{sec_optimization_procedure} we discuss the details  of steps (ii) and (iv).\\

Since the dynamical reduction hypothesis has been extensively demonstrated to  hold exactly for two-levels atoms through integrability~\cite{PhysRevLett.96.097005,PhysRevA.91.033628,PhysRevB.99.054520}, we focus on the $N=3$ levels case.
In Fig.~\ref{fig_fit_phaseIII_in_SU3} we show the results obtained in the spin-exchange model at $W/(\chi\mathcal{N}_a)=0.1$ by simulating the full many-body dynamics given by Eq.~\eqref{eq_SUN_W_neq_0} (continuous line). We initialize the system in a permutation invariant coherent state (cf.  Eq.~\eqref{eq_bosonic_coherent_state}) which displays Phase-III at $W=0$, and we consider photon-matter couplings $g_1 \neq g_2$, so that $\chi_{1,1}\neq \chi_{2,2}\neq \chi_{1,2}$ in Eq.~\eqref{eq_SUN_W_neq_0}.  
The black dashed lines are obtained from the numerically optimized single-body effective Hamiltonian. The dynamics of collective observables obtained via the effective Hamiltonian match well the dynamics obtained via the full many-body mean field Hamiltonian; this suggests  not only that the dynamical response observed in the homogeneous case is robust, but also that the dynamical reduction hypothesis holds at finite $W$. 

Due to   exponential sensitivity to initial conditions the procedure described above does not converge with high enough accuracy in the $SU(3)$ system in the chaotic phase of Sec.~\ref{sec_chaos_induced_fluctuactions}, which however persists also for weak inhomogenities.

\begin{figure}[t!]
\includegraphics[width=\linewidth]{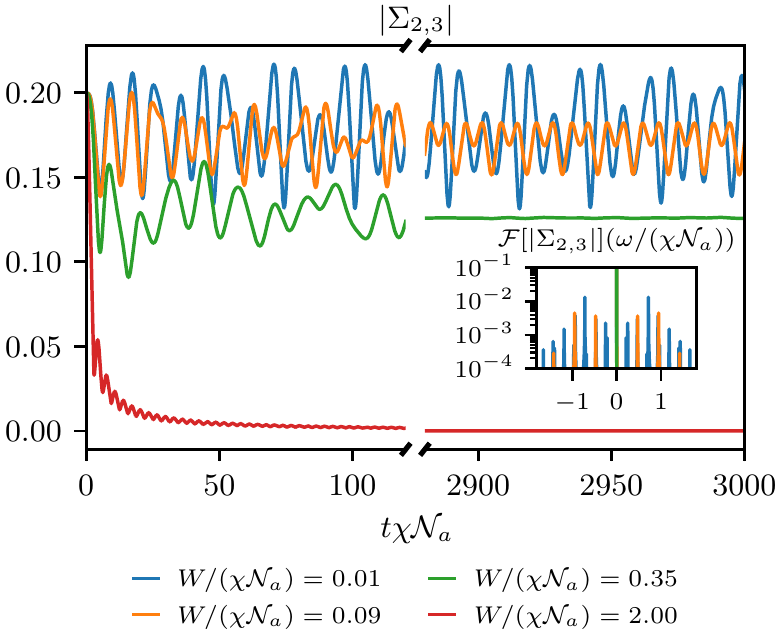}
\caption{Dynamics of the magnitude of the average phase coherence $|\Sigma_{2,3}(t)|$ in the $N=4$ levels spin-exchange model for $\chi_{1,2}\neq \chi_{2,3}\neq \chi_{3,4}$. The initial state is a permutation invariant (in space) coherent state.
As $W/(\chi\mathcal{N}_a)$ increases, 
the system displays different dynamical responses, passing from Phase-IV to Phase-III, then to Phase-II and eventually to Phase-I. The different phases are separated by crossover  regions where the dynamical responses cannot be sharply identified (not shown here). In the inset we show the {magnitude of the} Fourier spectrum in the late time dynamics. 
The results shown are obtained with $L=10^4$ sites and are not appreciably affected upon increasing $L$.
}
\label{fig_SU4_swiping_W}
\end{figure}

\begin{figure}[t!]
    \centering
    \begin{center}
    $W/(\chi\mathcal{N}_a)=0.08$
    \end{center}
        \includegraphics[width=0.46\linewidth]{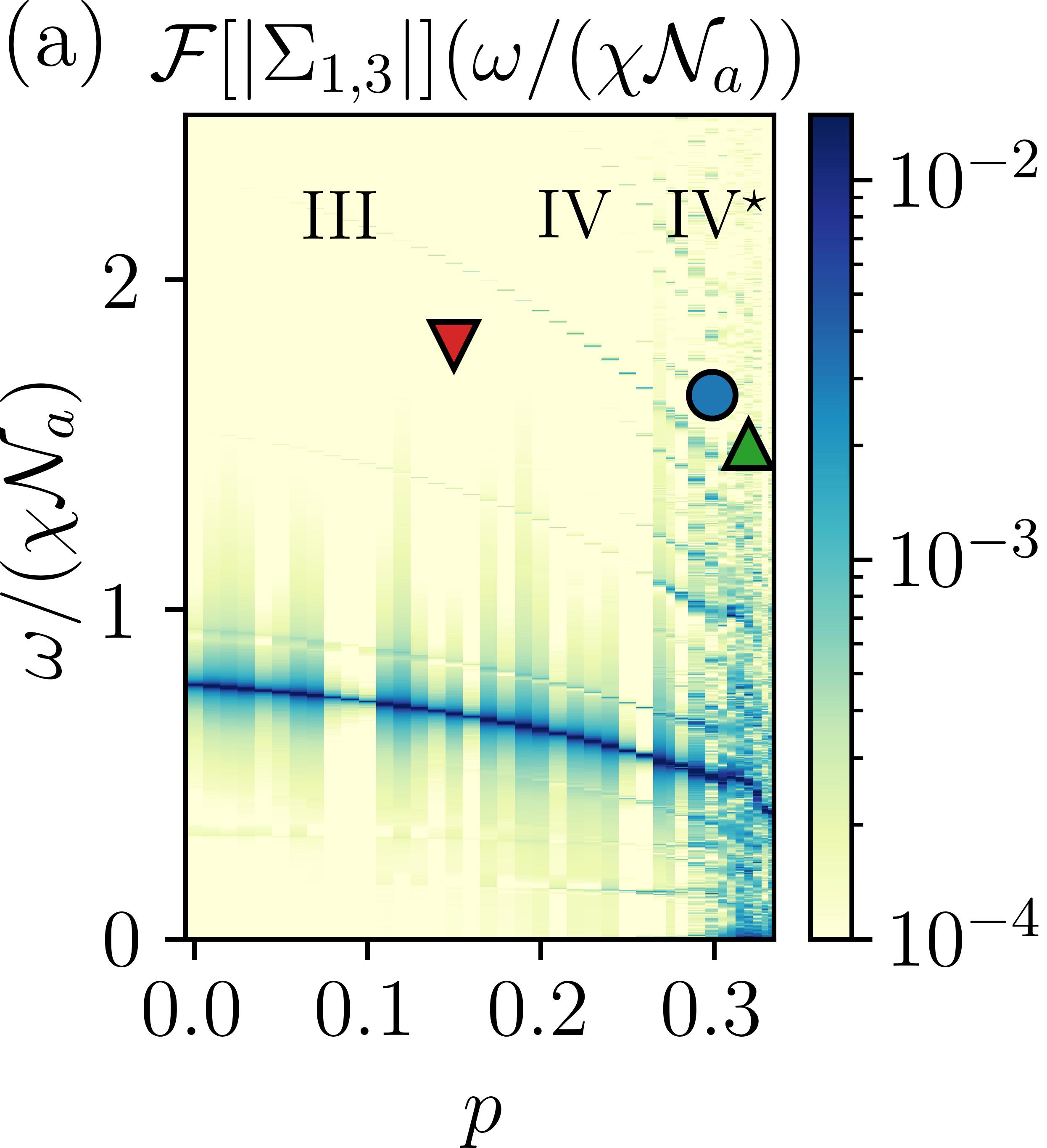}
                            \includegraphics[width=0.49\linewidth]{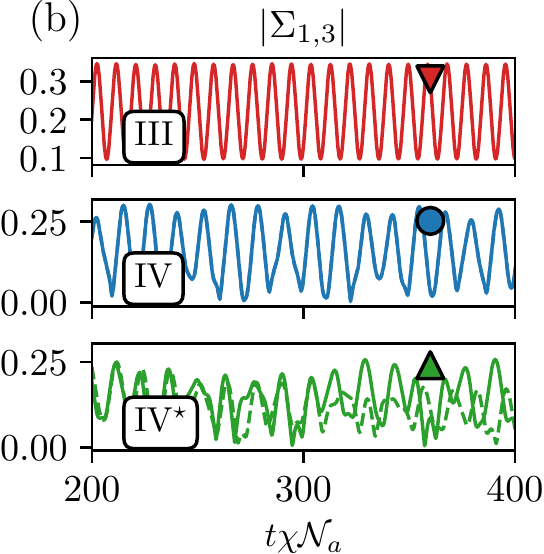}
                                \begin{center}
    $p=0.32$
    \end{center}
                    \includegraphics[width=0.46\linewidth]{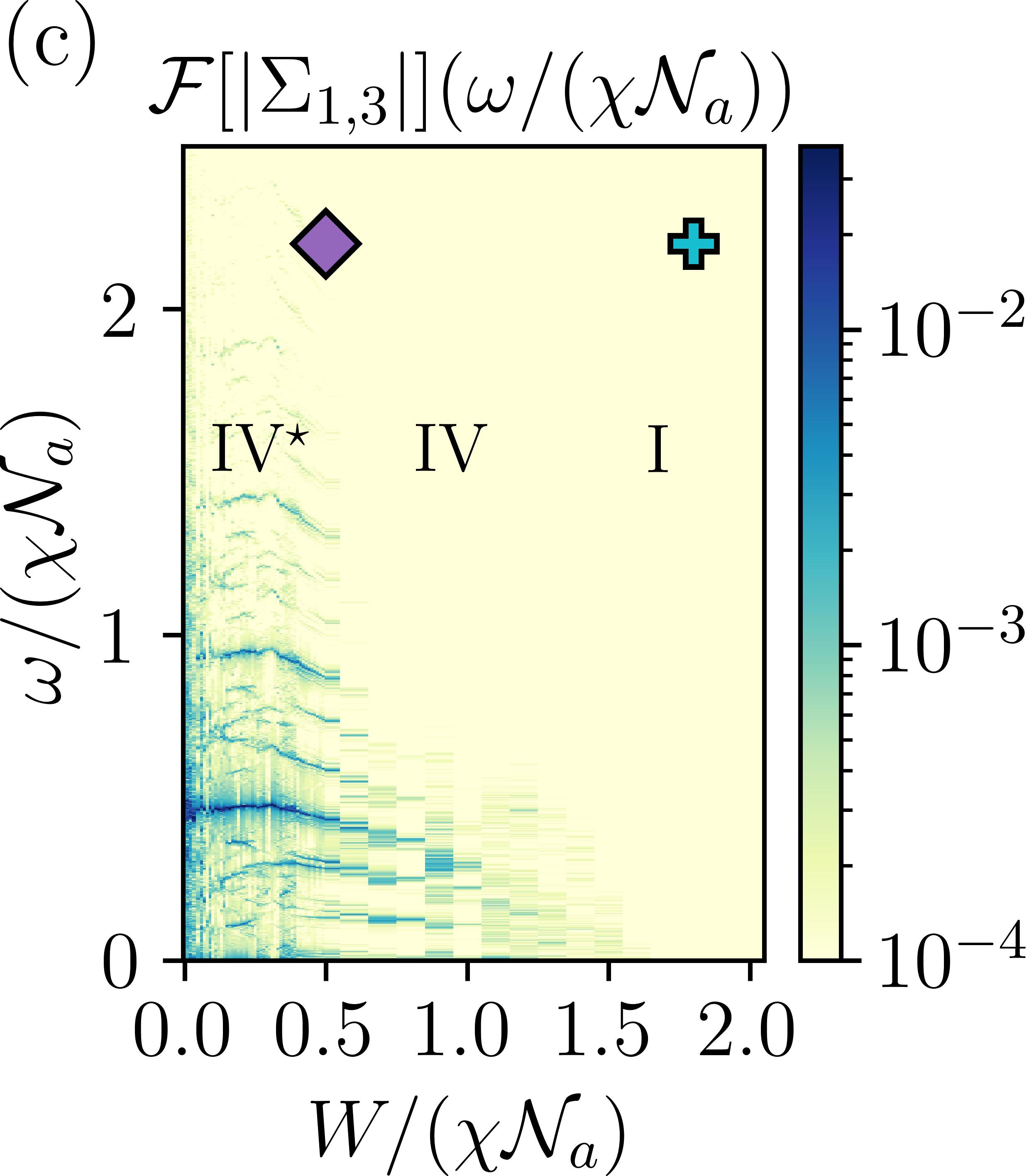}
               \includegraphics[width=0.49\linewidth]{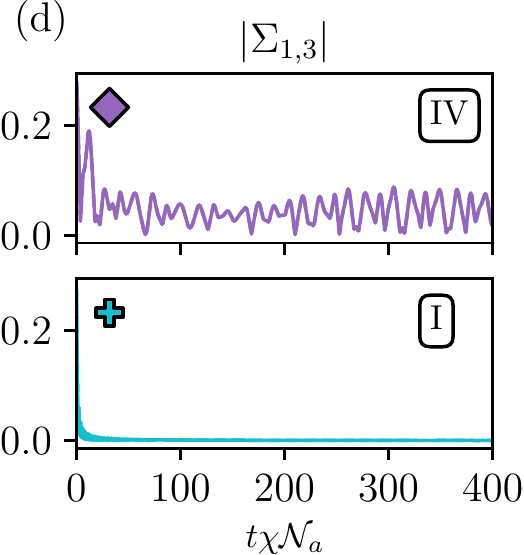}
    \caption{
    {Dynamical response in the $N=3$ levels spin-exchange model (Eq.~\eqref{eq_SUN_W_neq_0}) with inhomogeneity $W/(\chi\mathcal{N}_a)$ and $g_1/g_2 \approx 2$ ($\chi_{1,1}\neq \chi_{1,2} \neq \chi_{2,2}$), initializing a multimode Schr\"{o}dinger cat state parameterized by $p$ (cf. Eq.~\eqref{eq_parametrization_quantum_chaotic_state}) on each site. We focus on $|\Sigma_{1,3}|$; the other phase-coherences behave similarly. 
In the top panels we show the magnitude of the Fourier spectrum and dynamics of $|\Sigma_{1,3}|$, starting from two   close-by  initial states, changing $p$ at fixed $W/(\chi\mathcal{N}_a)=0.08$. 
Instead, in the bottom panels we change $W/(\chi\mathcal{N}_a)$ keeping $p=0.32$ fixed. 
As $p$ increases, and  $W/(\chi\mathcal{N}_a)$  is fixed,
we can infer from the Fourier spectrum a crossover from a regime with well-resolved peaks with commensurate frequencies (Phase-III), to multiple peaks with incommensurate frequencies (Phase-IV) and eventually to spectrum with multiple broad peaks typical of chaotic dynamics (Phase-IV${}^\star$). In (b) we show three examples of the different dynamical responses at $p=\{0.15,0.299,0.32\}$ (marked in the plot of the Fourier spectrum). 
As $W/(\chi\mathcal{N}_a)$ increases the system passes from Phase-IV${}^\star$ to Phase-IV and eventually Phase-I. In the (d) panel we show the dynamics at $W/(\chi\mathcal{N}_a)=\{0.5,1.8\}$, corresponding to Phase-IV and I (marked in the plot of the Fourier spectrum), respectively. The results shown are obtained with $L=10^4$ sites and are not appreciably affected upon increasing $L$.}
    }

%
    \label{fig_dynamical_response_SU3}
\end{figure}

\subsection{Robustness of dynamical responses to inhomogeneities \label{sec_dpt_via_W}}
Inhomogeneities in the local fields are generally expected to have an impact on the dynamics of collective observables.
One could argue that for $W/(\chi\mathcal{N}_a) \gg 1$ the local fields $\{h_n^{(j)}\}$   will dominate dynamics, and   phase-coherences would be washed out~(Phase-I). Here, we explore the dynamical responses at moderate inhomogeneity in spin-exchange Hamiltonian of Eq.~\eqref{eq_SUN_W_neq_0} for $N=3$ and ${N=}4$ levels atoms. In the $N=3$ levels case we initialize a multimode Schr\"{o}dinger cat state on each site for which the dynamical response is chaotic (Phase-IV${}^\star$) at $W=0$. On the contrary,  in the $N=4$ levels case we consider a coherent state for which the dynamical response is aperiodic (Phase-IV) at $W=0$. In both cases, $|\Sigma_{n,m}|$ displays dynamical responses different from Phase-I and Phase-II (relaxation) for $W/(\chi\mathcal{N}_a)\lesssim 1$.

First, let us consider the {$N=4$ } spin-exchange model in Eq.~\eqref{eq_SUN_W_neq_0}. In Fig.~\ref{fig_SU4_swiping_W} we show the dynamics of $|\Sigma_{2,3}|$ for different values of inhomogeneity $W/(\chi\mathcal{N}_a)$. We fix as initial condition a permutation invariant (in space) coherent state with different amplitudes on each level. The intra-level phase coherences $|\Sigma_{n,m}|$ displays Phase-IV up to a finite value of inhomogeneity $W/(\chi\mathcal{N}_a)$. Upon increasing $W/(\chi\mathcal{N}_a)$, $|\Sigma_{n,m}|$ displays Phase-III, Phase-II and eventually Phase-I. The different dynamical responses are divided by regions (not shown in Fig.~\ref{fig_SU4_swiping_W}) where the distinction between the different dynamical responses becomes more blurry.

Similarly, in the $SU(3)$ case, the dynamical response  generally passes from Phase-$Y$ to Phase-$(Y-1)$ as $W/(\chi\mathcal{N}_a)$ is increased (at fixed initial state), and with the dynamics eventually entering Phase-I due to the dominant inhomogeneous local fields {(bottom panels in Fig.~\ref{fig_dynamical_response_SU3})}. Instead, as initial quantum correlations in the initial state increases with $p$, the dynamical response generally passes from Phase-$Y$ to Phase-$(Y+1)$ {(top panels in Fig.~\ref{fig_dynamical_response_SU3}).
Additionally, we highlight that the dynamical responses are generally robust against small breaking of the permutation symmetry in the initial state}.

The robustness of the various  dynamical responses   against inhomogeneous local fields for $W>0$ can be ascribed to the many-body gap $\propto \chi\mathcal{N}_a$ that suppresses local spin flips and favors spin alignment. This mechanism has been shown to protect phase   coherence  
  in the spin-exchange {model} between $SU(2)$ spins~\cite{Norcia2018,PhysRevA.77.052305,PhysRevLett.125.060402}, and it is likely present also in our $N$ levels case.
As $W$ increases, this `many-body gap protection' is less effective and dephasing processes between the $SU(N)$ takes over. As a result, within the framework of  the dynamical reduction hypothesis, the number of effective sites  required to describe dynamics is reduced and accordingly the dynamical responses change.

For instance, in  the $SU(4)$ case (cf. Fig.~\ref{fig_SU4_swiping_W}), at   moderate inhomogeneity the system has $M>2Q$ effective DOFs, which lead to to Phase-IV observed in the $W=0$ case. As $W$ becomes sizeable,  dephasing starts to affect dynamics  and, since $M \leq 2Q$, the system displays  Phase-III, II and eventually I,  upon increasing the degree of inhomogeneity. The effects of dephasing   are apparent in the Fourier spectrum of $|\Sigma_{n,m}(t)|$ (cf. inset of Fig.~\ref{fig_SU4_swiping_W}), where the various Fourier components are `depleted' as $W$ increases until the whole spectrum becomes flat in Phase-I.  Similarly in the $SU(3)$ case, inhomogeneity leads to a loss of effective phase-space, a reduction of the effective DOFs and correspondingly { leads to a} loss of chaos~({the Lyapunov exponent} vanishes). {Along the same argument, the number of effective DOFs $M$ increases as initial correlations in the initial state increase, thus the system could enter in a regime with different dynamical responses  and eventually display chaotic behavior, as observed in the homogeneous case of Fig.~\ref{fig_classical_fluctuactions_chaos}.}
We highlight that the dynamical response displayed does not necessarily have to pass smoothly  from Phase-$Y$ to Phase-$(Y \pm 1)$, but there can be a `jump', as in the $SU(3)$ case where Phase-IV turns into Phase-I (see Fig.~\ref{fig_dynamical_response_SU3}), without displaying Phase-III and Phase-II. This has been also reported in the   integrable $SU(2)$ case~\cite{PhysRevLett.126.173601}.

\section{Experimental implementation \label{sec_experimental_implementation}}
\begin{figure}[t!]
\centering
\includegraphics[width=\linewidth]{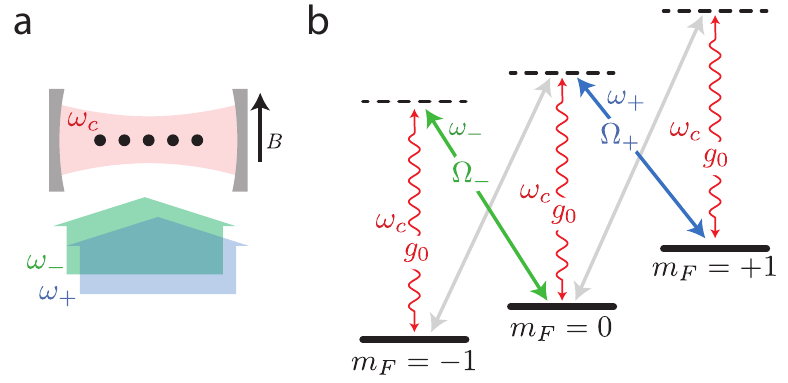}
\caption{Suggested experimental implementation. (a) Ensembles of $3$-level atoms in $L$ traps (black dots) are coupled to a single mode of an optical cavity (red) and transversely illuminated by a two-frequency laser field (blue, green).  (b) The applied magnetic field $B$ leads to a non-degenerate atomic level splitting allowing to selectively drive cavity-assisted Raman transitions. Coupling via the transverse laser fields (the cavity) are shown as solid (wiggly) arrows. If additional frequency components (grey arrows) are introduced, co- and counter-rotating terms can be engineered. \label{fig:levelscheme}}
\end{figure}

A possible experimental scheme to implement the couplings of Hamiltonian Eq.~(\ref{eq_H_sun_adiabatic}) is sketched in Fig.~\ref{fig:levelscheme}. Ensembles of $\mathcal{N}_a/L$ atoms are trapped at $L$ fixed positions and collectively coupled to a single mode of a high finesse optical cavity with resonance frequency $\omega_c$. At the same time, the atoms are subject to a multi-frequency laser field. The atoms are assumed to have a manifold of ground state sublevels which can be coupled using Raman transitions. If one leg of such a Raman transition is driven by a classical field while the second leg is coupled to the resonator mode, cavity-assisted Raman transitions can be implemented~\cite{Zhiqiang:17,PhysRevLett.122.010405,PhysRevX.11.041046}. In a microscopic description, a photon from the laser field is scattered into the cavity, while the internal state of an atom in one of the ensembles is changed. The photon is delocalized over the cavity mode and can subsequently drive a second Raman transition in another atomic ensemble. In this process, the photon is absorbed by an atom and then emitted into the driving laser field via bosonic stimulation.

As example we consider $^{87}$Rb atoms, where the $F=1$ ground-state hyperfine manifold has $N=3$ magnetic sublevels $m_F=(0,\pm 1)$. A sufficiently strong applied magnetic field leads to a non-degenerate level splitting due to linear and quadratic Zeeman shifts,  as sketched in the figure. In combination with a two-frequency transverse laser field, this allows to drive state-selective, cavity assisted two-photon Raman transitions between these states, as indicated in Fig.~\ref{fig:levelscheme}b. The laser field is far detuned from atomic resonance to avoid any spontaneous decay of the excited atomic state.
The frequencies $\omega_+$ and $\omega_-$ of the transverse laser field are chosen to match the different atomic level splittings when a photon is absorbed from or emitted into the cavity mode. For example, an atom at a specific site can be transferred from $m_F=0$ to $m_F=-1$ by absorbing a photon from the laser field at frequency $\omega_-$ and emitting a photon into the cavity. The very same cavity photon can then drive a transition at a different site where an atom in $m_F=0$ absorbs that photon and undergoes a transition to $m_F=+1$ while emitting into the laser field at frequency $\omega_+$. These two-photon Raman transitions correspond to the processes proportional to $g_n$ in Hamiltonian Eq.~(\ref{eq_H_sun_adiabatic}), where the coupling strengths $g_n$ can be engineered via the single-photon Rabi frequencies $\Omega_+$ and $\Omega_-$.

The co-rotating terms proportional to $\lambda_n$ in Hamiltonian Eq.~(\ref{eq_H_sun_adiabatic}) can be implemented if additional laser frequencies are added to the transverse laser field. Such couplings are indicated by the grey arrows in Fig.~\ref{fig:levelscheme}b. The relative strengths of the co- and counterrotating terms can be independently tuned via the respective Rabi rates of the driving laser fields~\cite{PhysRevX.11.041046}.

This scheme can further be extended to $N>3$ by choosing atomic states with larger magnetic sublevel manifolds as they can be found for example in lanthanide atoms. Finally, site dependent energies $h^{(j)}_n$ of the atomic modes can be introduced by applying a magnetic field gradient along the cavity axis in addition to the homogeneous magnetic field~\cite{PhysRevLett.123.130601,Periwal2021}. In a realistic scenario also cavity decay due to losses  at the mirrors has to be taken into account. Its influence can however be reduced by introducing a detuning between the cavity resonance and the frequency of the field scattered into the cavity mode.
\section{Discussion \label{sec_discussion}}

\subsection{Role of dissipation \label{sec_role_dissipation}}
In our analysis  we have considered the system completely isolated from the environment.
In cavity-QED systems there are two main sources of dissipation,  free-space emission of single-atom excitations  and loss of the cavity field. Let us denote the rates of the two processes with $\eta$ and $\kappa$, respectively, and their  jump operators with $\hat{L}_n^{(j)} = \sqrt{\eta}\hat{\Sigma}_{n,n+1}^{(j)}$ and $\hat{L} = \sqrt{\kappa}\hat{a}$, where   $j\in[1,L]$ and $n \in [1,N-1]$. 
The relevant time scales for the coherent dynamics are set by the collective photon-matter couplings $\lambda_n\sqrt{\mathcal{N}_a}$ and $g_n\sqrt{\mathcal{N}_a}$.
The different dynamical responses can be dominantly  ascribed to Hamiltonian dynamics if  $\lambda_n\sqrt{\mathcal{N}_a},g_n\sqrt{\mathcal{N}_a} \gg \kappa,\eta$. 

We   provide a more accurate estimate    in the far detuned cavity mode regime, where all the results of this work have been derived. 
Focusing on the $SU(N)$ spin-exchange case for simplicity , the photon effectively induces elastic all-to-all interactions of strength  $\chi_{n,m}=g_n g_m \omega_0/((\omega_0^2+(\kappa/2)^2)$; in addition,   the collective atomic transitions are radiatively broadened by the coupling to the cavity, leading to collective decays with rate per-particle $\Gamma_n = g_n^2 \kappa/(\omega_0^2 + (\kappa/2)^2)$~\cite{2008,PhysRevLett.129.063601,PhysRevA.99.033845,PhysRevLett.116.153002,PhysRevA.95.063852}. 
The coherent dynamics are fast with respect to the time scales of the dissipation if $\mathcal{N}_a\chi_{n,m} \gg \{\mathcal{N}_a\sqrt{\Gamma_{n}\Gamma_m},\eta\}$, which translate to $\omega_0 \gg \kappa$ and $\mathcal{N}_a\chi_{n,m} \gg \eta$ (see Appendix~\ref{appendix_effect_of_losses_details} for the complete derivation). In this parameters' regime, dynamics are basically ruled only by coherent evolution, at least up to times parametrically large in  $\omega_0/\kappa$ and $\mathcal{N}_a\chi_{n,m}/\eta$.
\subsection{Connection with $SU(N)$ fermionic systems \label{sec_extension_results_to_other_systems}}
As already anticipated in Sec.~\ref{sec_initial_states},   the number of atoms per site $\mathcal{N}_a/L$ is a conserved quantity in our system and our results can be extended to a large class of systems which can be mapped to the Hamiltonians in Eqs.~\eqref{eq_H_SUn_wphoton} and~\eqref{eq_H_sun_adiabatic}. 
As an example, let us consider a $N$-level fermionic system with annihilation (creation) operators $\hat{c}_{n,j}^{(\dagger)}$ with $n\in [1,N]$ and site index $j$. We can define the pseudospins $\hat{\Sigma}_{n,m}^{(j)} \equiv \hat{c}_{n,j}^\dagger \hat{c}_{m,j}$~\cite{auerbach2012interacting}, which in turn can be expresses in terms of Schwinger bosons as $\hat{\Sigma}_{n,m}^{(j)} = \hat{b}_{n,j}^\dagger \hat{b}_{m,j}$. If via this procedure the fermionic Hamiltonian as a function of the Schwinger bosons is identical to one of the Hamiltonians here investigated, our results obtained via coherent states could be applied straightforwardly. 
Indeed, the mean field at the level of the Schwinger bosons is mathematically equivalent to simulating pure single-particle states $|\psi_j\rangle = \sum_{n=1}^N \langle \hat{b}_{n,j}\rangle|n_j\rangle$, with the caveat of interpreting the bosonic amplitudes as probability amplitudes~\cite{perlin2022engineering}.
\subsection{Roadway towards a universal dynamical reduction hypothesis}
In this work we have formulated and tested a  reduction hypothesis for the dynamics of $SU(N)$   cavity QED systems with atoms in a multilevel ladder configuration. 
We found that the reduction hypothesis was a useful description for a variety of systems with different initial states, levels, inhomogeneous fields, and light matter interactions.
This plethora of applications calls naturally for a broader framework. 
Our classification of dynamical responses based on the dynamical reduction hypothesis and the Arnold-Liouville theorem might posses  the flavor of universality. 
It would be in fact extremely interesting to encompass all the specific examples mentioned above, under the   lenses of the  symmetries both of   the local DOFs and of the   light-matter interactions.
 Similarly to   renormalization group approaches, one could explore   the basins of attraction of the effective few-body models presented here.   
 Upon changing the symmetries and the   conservation laws of a given macroscopic model, one could expect to  distinguish a set of irrelevant perturbations in which the reduction hypothesis remains valid, and set of relevant perturbations in which the reduction hypothesis fails and no effective few-body model describes the dynamics of collective observables.

 From our numerical experiments it seems natural that perturbations that do not dramatically change the long-range nature of the interactions would be irrelevant.
 We would therefore expect similar dynamical responses in the presence of   different photon-assisted transitions (e.g.   from the $n$-th level to any $m$-th level level via a single-photon process), or of squeezed terms (e.g. $\propto (b^\dagger b^\dagger a + h.c.)$), where the information about the state cannot be retrieved solely by the $SU(N)$ coherences, but would also require  terms such as $\langle b^\dagger b^\dagger\rangle$~\cite{Marino2019}.
 Investigations into the latter perturbations might disclose connections between    multi-mode squeezing and the generation of universal   dynamical responses.
It is also completely natural to investigate the impact of different level configurations, for instance, studying the case of two degenerate subspaces of excited and ground states~\cite{PhysRevLett.128.153601,PhysRevX.12.011054,chu2022photon}.
A preliminary analysis suggests that  they are also irrelevant perturbations and that a reduction hypothesis still holds here.
Finally we note that $SU(N)$ generalizations of BCS models~\cite{PhysRevA.104.043316,PhysRevA.85.041604,PhysRevLett.109.205305,PhysRevA.89.043610} would, under a generalized Anderson pseudo-spin mapping, have a similar form as to the models we study here and also be describable by a reduction hypothesis. 

In contrast, any perturbation that introduces short range interactions could be expected to be relevant to the effective few-body Hamiltonian basin of attraction.
This appears to be the case in the context of time crystals~\cite{seetharam2022correlation,PhysRevLett.129.050603,PhysRevB.105.184305,10.21468/SciPostPhysCore.4.3.021,PhysRevB.104.134309,https://doi.org/10.48550/arxiv.2208.11659}, where short range interactions generally melt the time crystal at late times and lead to generically asynchronous relaxation.
Separability of the interactions will also likely play a role: systems with separable interactions seem describable by an effective few-body model~\cite{kelly2022resonant,PhysRevX.9.041011}; while models with inseparable interactions can lead to glassy relaxation~\cite{PhysRevLett.107.277202,PhysRevLett.107.277201,Kelly2020} and cannot be described by effective few-body models~\cite{PhysRevA.87.063622,PhysRevX.9.041011}.
Furthermore, systems with a number of atomic levels comparable to the number of sites, $N\sim L$, may also pose obstacles in defining    an effective few-body theory, but could be relevant for experiments in synthetic dimensions~\cite{https://doi.org/10.48550/arxiv.2204.06421,https://doi.org/10.48550/arxiv.2208.01896}.
Naturally the effects of dissipation would also not be captured by a few-body Hamiltonian picture, but instead potentially be described by a few-body dissipative model such as a Lindbland master equation. 

The strong numerical-oriented approach we have taken here has provided serious evidence of a description using an effective few-body model, even demonstrating a near perfect ability to capture the dynamics of collective observables.
Still, an analytic approach could yield important insights and provide a more solid ground for classifying different perturbations as relevant or irrelevant to the few-body attractive basin.
 Considering the variety of AMO systems modeled by collective interactions, finding such description would constitute  a significant step forward in understanding universality out of equilibrium~\cite{Marino2022,PhysRevLett.123.230604}.  
\section*{Acknowledgements}
We thank E. Altman, M. Foster, A. Hemmerich, H. Kessler, A. Polkovnikov, A. M. Rey,  M. Schleier-Smith,  D. Stamper-Kurn, {M. Stefanini} for stimulating discussions. 
This project has
been supported by the Deutsche Forschungsgemeinschaft (DFG, German Research Foundation) through the
Project ID 429529648-TRR 306 QuCoLiMa (``Quantum Cooperativity of Light and Matter''), 422213477 - TRR 288 (project
B09), and the grant
HADEQUAM-MA7003/3-1; by the Dynamics and
Topology Center, funded by the State of Rhineland Palatinate; and in part by the National Science Foundation under Grant No. NSF PHY-1748958 (KITP program `Non-
Equilibrium Universality: From Classical to Quantum and Back'). We further acknowledge funding from the SNF (project number IZBRZ2 186312). The work of R.F. has been supported by the ERC under grant  agreement n.101053159 (RAVE). Parts of this research were conducted using the Mogon supercomputer and/or advisory services offered by
Johannes Gutenberg University Mainz (\url{hpc.uni-mainz.de}), which is a
member of the AHRP (Alliance for High Performance Computing in Rhineland Palatinate,  \url{www.ahrp.info}), and the Gauss
Alliance e.V. We gratefully acknowledge the computing time granted on the Mogon supercomputer at Johannes Gutenberg University Mainz (\url{hpc.uni-mainz.de}) through the project ``DysQCorr.''
\newpage
\appendix
\section{Computation of the Lyapunov exponent \label{appendix_computation_lyapunov}}
\begin{figure}[b!]
\centering
\includegraphics[width=\linewidth]{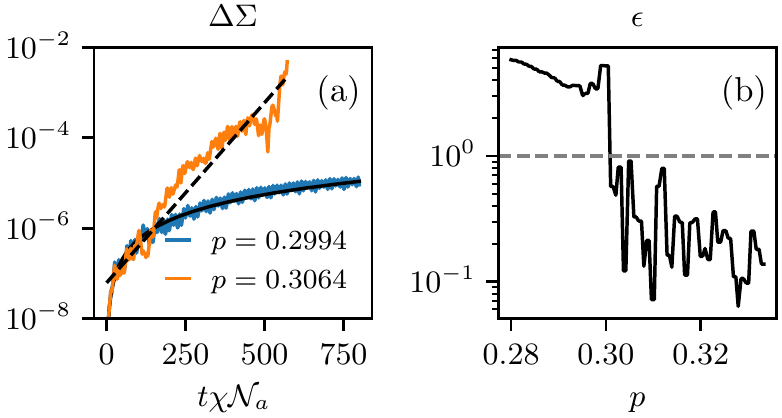}
\caption{(a): Dynamics of the average distance between $R=12$ nearly sampled trajectories starting from two multimode Schr\"{o}dinger cat states parameterized by  $p$ (cf. Eq.~\eqref{eq_parametrization_quantum_chaotic_state}) {in the $N=3$ levels spin-exchange model}. We fix two different values of $p=\{0.2994,0.3064\}$. The continuous line is the polynomial fit. The dashed line is the exponential fit. (b): Relative error between the exponential fit and the polynomial fit. When $\epsilon>1$ ($\epsilon<1$) the polynomial (exponential) fit better approximates the data. The horizontal dashed gray line is at $\epsilon=1$. Notice the sharp change of $\epsilon$ around $p\approx 0.3$. The parameters of the simulations are the same as the one in Fig.~\ref{fig_classical_fluctuactions_chaos} in the main text.}
\label{fig_lyapunov_exponent_calculation_details}
\end{figure}
Here we give further details about the calculation of the Lyapunov exponent referring specifically to the results in Sec.~\ref{sec_chaos_induced_fluctuactions}. 
We extract the Lyapunov exponent investigating the divergence of $R$ nearly sampled initial conditions. We use as measure of the distance of two trajectories the Frobenious norm of the difference of the average one-body reduced density matrices, namely 
\begin{equation}
\Delta\Sigma(i,j,t) =  \sqrt{\sum_{n,m=1}^N \left|\Sigma_{n,m}(i,t) - \Sigma_{n,m}(j,t)\right|^2}
\end{equation}
where $i,j\in[1,R]$ label the trajectory and $t$ is the time. The $R$ initial states are sampled such that $\Delta\Sigma(i,j\neq i,t=0) \approx 10^{-8}$. Then, we compute the average distance over all the trajectories
\begin{equation}
\Delta\Sigma(t) = \frac{2}{R(R-1)}\sum_{i=1}^{R} \sum_{j=i+1}^R \Delta\Sigma(i,j,t).
\end{equation}
The dynamics is regular when $\Delta\Sigma(t)$ grows polynomially in time, while it is chaotic if $\Delta\Sigma(t)$ grows exponentially in time, with the largest Lyapunov exponent equal to the rate of the exponential. In Fig.~\ref{fig_lyapunov_exponent_calculation_details}(a)
we show two paradigmatic examples in the regular phase and chaotic phase. Specifically, referring to the results in Fig.~\ref{fig_classical_fluctuactions_chaos}, we fix $p \lesssim p^\star$ and $p\gtrsim p^\star$ with $p^\star\approx 0.3$, in order to highlight the abrupt change of the behavior of $\Delta\Sigma(t)$. In Fig.~\ref{fig_lyapunov_exponent_calculation_details}(b) we show the relative error $\epsilon$ between the exponential fit and the polynomial fit. When $\epsilon>1$ ($\epsilon<1$) the polynomial (exponential) fit better approximates the data. We checked that our results are not affected by decreasing the time step. 

\section{Chaos induced upon trading $SU(2)$ with $SU(3)$ interactions in three-level system \label{appendix_dynamical_response_chaos_SU3}}
\begin{figure}[b!]
\centering
\includegraphics[width=\linewidth]{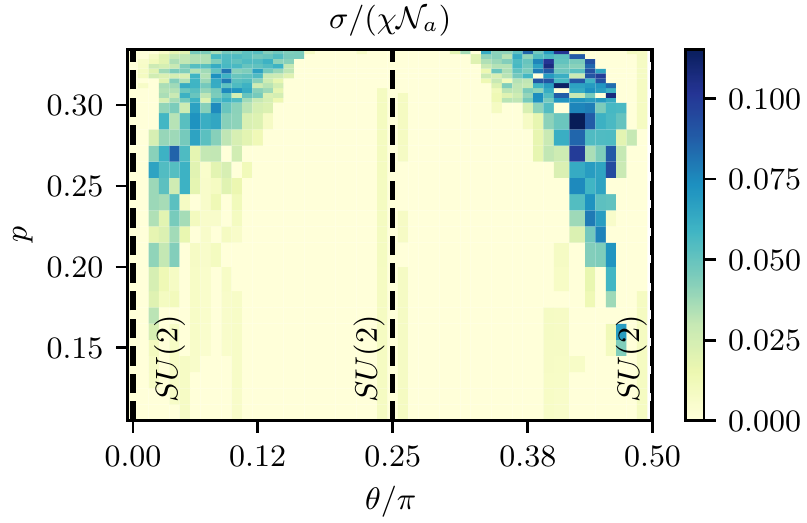}
\caption{Lyapunov exponent as a function of $p$ and $\theta$ in the {$N=3$ } spin-exchange model. The initial state is a multimode Schr\"{o}dinger cat state parameterized by $p$ via Eq.~\eqref{eq_parametrization_quantum_chaotic_state}.
We set $\mathcal{N}_a\to\infty$ and $W=0$. The vertical dashed lines are along the $SU(2)$ limit, where the dynamics is effectively taking place in a $SU(2)$ subgroup of $SU(3)$. The all-to-all couplings are parameterized such that $\chi_{1,1} < \chi_{2,2}$ for $\theta/\pi \in [0,0.25)$, $\chi_{1,1} > \chi_{2,2}$ for $\theta/\pi \in (0.25,0.5]$ and $\chi_{1,1}=\chi_{2,2}$ at $\theta/\pi=0.25$.
In the $SU(2)$ limits ($\theta/\pi = \{0,0.25,0.5\}$) the dynamical response is regular for any value of $p$, since the couplings constrict the dynamics to take place in a $SU(2)$ subspace of $SU(3)$. Deep in the $SU(3)$ limit ($\theta/\pi \approx 0.1$ and $\theta/\pi \approx 0.4$), $|\Sigma_{n,m}|$ displays exponential sensitivity for any value of $p \gtrsim p^\star$, with $p^\star$ dependent on $\theta$. Near the $SU(2)$ two-level limits ($\theta/\pi=\{0,0.5\}$) we observe chaotic regions embedded in regular ones.}
\label{fig_appendix_lyapunov_exponent_p_vs_theta}
\end{figure}

Here, we investigate the onset of a chaotic phase in the three-level spin-exchange model starting from a separable state in Eq.~\eqref{eq_generic_state} with $|\psi_j\rangle$ in a Schr\"{o}dinger cat state parameterized by $p$ via Eq.~\eqref{eq_parametrization_quantum_chaotic_state}. In Fig.~\ref{fig_appendix_lyapunov_exponent_p_vs_theta} we show the maximum Lyapunov exponent as a function of the ratio $g_1/g_2 = \tan(\theta)$ and $p$ in the homogeneous case ($W=0$). For $\theta/\pi = \{0,0.25,0.5\}$ the Hamiltonian can be written in terms  of a $SU(2)$ subgroup of $SU(3)$, and since   dynamics are therefore  restricted to that subgroup there is no chaos for any value of $p$. Furthermore, for $\theta/\pi=\{0,0.5\}$ we have $g_1=0$ and $g_2=0$, respectively, thus we recover the $SU(2)$ two-level system limit. As $\theta/\pi$ deviates from the $SU(2)$ limits, the system displays chaotic behavior for sufficiently large $p$. Deep in the $SU(3)$ limit, we observe chaos for any $p \gtrsim p^\star$, while near the $SU(2)$ limit there are islands of chaotic behavior embedded in regular ones. We compute the Lyapunov exponent following the procedure described in Appendix~\ref{appendix_computation_lyapunov}. Additionally, we manually set $\sigma/(\chi\mathcal{N})=0$ when it is less than $0.01$, since our procedure was signaling chaos in regions where, by direct inspection, there were no signatures of it.

\section{Chaos induced by a finite fraction of Schr\"{o}dinger cat states in $SU(3)$ spin-exchange Hamiltonian \label{appendix_finite_fraction_chaos}}

\begin{figure}[t!]
\centering
\includegraphics[width=\linewidth]{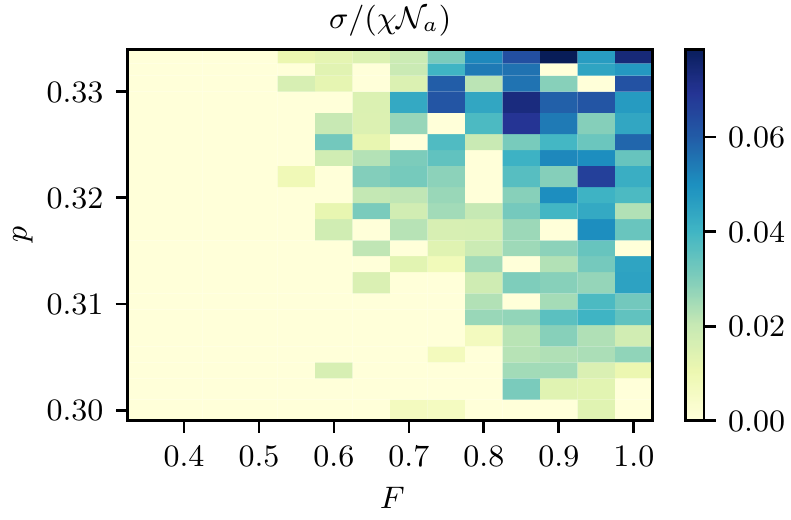}
\caption{Lyapunov exponent for the $N=3$ levels spin-exchange Hamiltonian in the homogeneous case ($W=0$) as a function of the initialized fraction $F$ of multimode Schr\"{o}dinger cat states in the state in Eq.~\eqref{eq_psi_cat_coherent}. 
The results were obtained simulating $L=100$ sites at fixed $g_1/g_2\approx 2$ (in the units adopted in Fig.~\ref{fig_appendix_lyapunov_exponent_p_vs_theta} it corresponds to $\theta/\pi=0.36$). The results are not affected upon increasing $L$.}
\label{fig_chaotic_seed}
\end{figure}

Here, we provide additional details about the results discussed in Sec.~\ref{sec_chaotic_seed}. We consider the Schr\"{o}dinger cat state $|\psi_\text{cat}\rangle \sim (|\widetilde{\gamma}^{(1)}\rangle + |\widetilde{\gamma}^{(2)}\rangle)$ defined in Eq.~\eqref{eq_parametrization_quantum_chaotic_state} and parameterized via $p$, while we consider  $|\widetilde{\gamma}^{(1)}\rangle$ as coherent state. We initialize a fraction $F$ of Schr\"{o}dinger cat states $|\psi_\text{cat}\rangle$, such that the initial state is
\begin{equation}\
\label{eq_psi_cat_coherent}
|\Psi\rangle = \otimes_{j=1}^{\lfloor FL\rfloor}|\psi_\text{cat}\rangle \otimes_{j=\lfloor FL\rfloor + 1}^L |\widetilde{\gamma}^{(1)}\rangle,
\end{equation}
where $\lfloor x\rfloor$ returns the least integer greater than or equal to $x$.
The evolution is governed by the $SU(3)$ spin-exchange Hamiltonian at $W=0$ and $g_1/g_2\approx 2$ (in units adopted in Fig.~\ref{fig_appendix_lyapunov_exponent_p_vs_theta} it corresponds to $\theta/\pi=0.36$).
In Fig.~\ref{fig_chaotic_seed} we show the Lyapunov exponent as a function of $p$ and $F$. For $F < F^\star$ the system displays regular dynamics, while for $F > F^\star$ the system enters in a chaotic regime. The Lyapunov exponent is computed following the same procedure discussed in Appendix~\ref{appendix_computation_lyapunov}.

\section{Optimization procedure \label{sec_optimization_procedure}}
Here we give further details about the practical implementation of steps (ii) and (iv) in the optimization procedure discussed in Sec.~\ref{sec_Wfinite}.
We set as initial guess for the initial state $\widetilde{\Sigma}_{n,m}(t=0) = \Sigma_{n,m}(t=\tau)$, with the time $\tau$ large enough, such that the initial transient dynamics is neglected. We set $\widetilde{h}_n=\sum_{j=1}^L h_j^{(n)}/L=0$ and $\widetilde{\chi}_{n,m}=\chi_{n,m}$, as initial guesses for the parameters. Throughout the procedure, we keep $\widetilde{\zeta}_{n,m}=\widetilde{\nu}_{n,m}=0$ since they control processes absent in the bare model. Then, we numerically optimize both the initial state $\{\widetilde{\Sigma}_{n,m}(t=0)\}$ and the parameters $\{\widetilde{h}_n,\widetilde{\chi}_{n,m}\}$ in order to minimize the cost function in Eq.~\eqref{eq_cost_function} in the main text.
The optimization procedure stops when the relative change between two consecutive iterations of the guessed solutions is less than $\approx {10^{-2}}$.\\
{We also test the convergence of the optimization procedure modifying the cost function. Specifically, we consider as cost function the average norm-2 distance
\begin{equation}
\label{eq_eq_cost_function_2}
\epsilon_2 = \frac{1}{T}\int_0^T\sqrt{\sum_{n,m=1}^N\left|\widetilde{\Sigma}_{n,m}(t) - \Sigma_{n,m}(t)\right|^2}\:dt,
\end{equation}
and compare with the one based on the norm-1 cost function in Eq.~\eqref{eq_cost_function}. We test the two procedures using the same parameters and initial state of Fig.~\ref{fig_fit_phaseIII_in_SU3} in the main text, namely $W/(\chi\mathcal{N}_a)=0.1$, $g_2/g_1 \approx 10^{-2}$ and a permutationally invariant (in space) bosonic coherent state.  Once the two procedures converged, we compare them computing the norm-1 distance between the optimized $\widetilde{\Sigma}(t)$ and $\Sigma(t)$ obtained from the full many-body dynamics. 
We obtain $\epsilon_1/\epsilon_2\approx 0.6$ 
(where $\epsilon_2$ is the norm-1 computed at the end of the optimization procedure based on the minimization of the norm-2 in Eq.~\eqref{eq_eq_cost_function_2}) showing a slight advantage of norm-1 over the norm-2 in the optimization procedure.}

\section{Dynamics with cavity losses \label{appendix_effect_of_losses_details}}
The dynamics of the matter-light system can be described by the master equation for the density matrix
\begin{equation}
\frac{d\hat{\rho}}{dt} = -i[\hat{H},\hat{\rho}] + \mathcal{L}_c[\hat{\rho}]+\mathcal{L}_a[\hat{\rho}].
\end{equation}
Here, $\hat{H}$ is the Hamiltonian in Eq.~\eqref{eq_H_SUn_wphoton}, where now the photon is an active DOF, and
\begin{equation}
\begin{split}
\mathcal{L}_c[\hat{\rho}] &= \frac{\kappa}{2}\left(2 \hat{a} \hat{\rho}\hat{a}^\dagger - \hat{a}^\dagger \hat{a} \hat{\rho} -  \hat{\rho} \hat{a}^\dagger \hat{a}\right),\\
\mathcal{L}_a[\hat{\rho}] &= \frac{\eta}{2}\sum_{j=1}^L \sum_{n=1}^{N-1}\Big(2 \hat{\Sigma}_{n,n+1}^{(j)}\hat{\rho}\hat{\Sigma}_{n+1,n}^{(j)} \\
&- \hat{\Sigma}_{n+1,n}^{(j)} \hat{\Sigma}_{n,n+1}^{(j)} \hat{\rho} - \hat{\rho} \hat{\Sigma}_{n+1,n}^{(j)} \hat{\Sigma}_{n,n+1}^{(j)} \Big)
\end{split}
\end{equation}
are the Lindbland terms that describe the cavity-photon loss with decay rate $\kappa$ and emission of single-atom excitation with rate $\eta$.
From now on we set $\lambda_n=0$ in the Hamiltonian $\hat{H}$ and we consider the far detuned regime of the cavity  mode as we are mainly interested in the spin-exchange interaction case.
We perform   adiabatic elimination such that~\cite{2008,PhysRevLett.129.063601,PhysRevA.99.033845,PhysRevLett.116.153002,PhysRevA.95.063852} 
\begin{equation}
\hat{a}(t) \approx -\sum_{n=1}^{N-1}\frac{ig_n \hat{\Sigma}_{n,n+1}(t)}{\left(i\omega_0 + \kappa/2\right)}.
\end{equation}
In this regime the dynamics of the density matrix of the matter DOFs $\hat{\rho}_m$ is given by the matter-only master equation
\begin{equation}	
\frac{d\hat{\rho}_m}{dt} = -i[\hat{H}_\text{ad},\hat{\rho}_m] + \mathcal{L}_\Gamma[\hat{\rho}_m]+\mathcal{L}_a[\hat{\rho}_m].
\end{equation}
Here, $\hat{H}_\text{ad}$ is the Hamiltonian given in Eq.~\eqref{eq_H_sun_adiabatic}, with $\nu_{n,m}=\zeta_{n,m}=0$ and all-to-all couplings per-particle
\begin{equation}
\label{eq_all_to_all}
\chi_{n,m} = \frac{g_n g_m \omega_0} {\omega_0^2 + (\kappa/2)^2}.
\end{equation}
The dissipative part $\mathcal{L}_\Gamma[\hat{\rho}_m]$ is given by
\begin{equation}
\begin{split}
\mathcal{L}_\Gamma[\hat{\rho}_m] = &\sum_{n,m=1}^{N-1} \frac{\sqrt{\Gamma_{n}\Gamma_{m}}}{2} \Big(2 \hat{\Sigma}_{n,n+1}\hat{\rho}_m \hat{\Sigma}_{m+1,m} +\\
&- \hat{\Sigma}_{n+1,n} \hat{\Sigma}_{m,m+1} \hat{\rho}_m - \hat{\rho}_m \hat{\Sigma}_{n+1,n} \hat{\Sigma}_{m,m+1}\Big),
\end{split}
\end{equation}
where
\begin{equation}
\label{eq_decay_rate}
\Gamma_{n} = \frac{g_n^2 \kappa}{\omega_0^2 + (\kappa/2)^2}, 
\end{equation}
is the decay rate per-particle of the collective DOFs. Imposing $\mathcal{N}_a\chi_{n,m} \gg \mathcal{N}_a\sqrt{\Gamma_n\Gamma_m}$ and $\mathcal{N}_a\chi_{n,m} \gg \eta$ we obtain the conditions discussed in the main text.

\bibliography{DPT_paper}
\end{document}